\documentclass[hyper]{JHEP3} 
\pdfoutput=1

\usepackage{array}
\usepackage{amssymb}
\usepackage{amsmath}
\usepackage{epsfig}

\newcommand{\bal}{\begin{align}}

\ifpdf
\usepackage{graphicx}
\else
\usepackage{graphicx}
\fi

\allowdisplaybreaks

\usepackage{multicol,bbm}

\include{symbolsJHEP}

\title{On the new physics reach of the decay mode \boldmath{\BdbKsll}}

\author{Ulrik Egede \\
	Imperial College London, London SW7~2AZ, United Kingdom\\
	E-mail: \email{u.egede@imperial.ac.uk}}
\author{Tobias Hurth\\
	Institute for Physics, Johannes Gutenberg-University, D-55099 Mainz, Germany \\
	E-mail: \email{tobias.hurth@cern.ch}}
\author{Joaquim Matias \\
	Universitat Aut\`onoma de Barcelona, 08193 Bellaterra, 
Barcelona, Spain\\
	E-mail: \email{matias@ifae.es}}
\author{Marc Ramon\\
	Universitat Aut\`onoma de Barcelona, 08193 Bellaterra, 
Barcelona, Spain\\
	E-mail: \email{mramon@ifae.es}}
\author{Will Reece \\
	CERN, Dept. of Physics, CH-1211 Geneva 23, Switzerland\\
	Imperial College London, London SW7~2AZ, United Kingdom\\
	E-mail: \email{will.reece@cern.ch}}

\preprint{
  UAB-FT 679\\
  MZ-TH/10-04\\
  IC/HEP/2010-02\\
  CERN-PH-TH/2010-032\\
 } 

 \abstract{We present a complete method to construct QCD-protected
   observables based on the exclusive 4-body $B$-meson decay
   $\BdbKsll$ in the low dilepton mass region. The core of the method
   is the requirement that the constructed quantities should fulfil
   the symmetries of the angular distribution. We have
   identified all symmetries of the angular distribution in the
   limit of massless leptons and explore: a new non-trivial relation
   between the coefficients of the angular distribution,
   the possibility to fully solve the system for the \Kstar
   amplitudes and the construction of non-trivial observables.
   
   We also present a phenomenological analysis of the new physics
   sensitivity of angular observables in the decay based on QCD
   factorisation.  We further analyse the \CP-conserving observables,
   \AT2, \AT3 and \AT4. They are practically free of theoretical
   uncertainties due to the soft form factors for the full range of
   dilepton masses rather than just at a single point as for
   \AFB. They also have a higher sensitivity to specific new physics
   scenarios compared to observables such as \AFB. Moreover, we
   critically examine the new physics reach of \CP-violating
   observables via a complete error analysis due to scale dependences,
   form factors and $\Lambda_{\rm QCD}/m_b$ corrections.  We have developed an
   ensemble method to evaluate the error on observables from
   $\Lambda_{\rm QCD}/m_b$ corrections. Finally, we explore the experimental
   prospects of \CP-violating observables and find
   that they are rather limited. Indeed, the \CP-conserving (averaged)
   observables $A_{\mathrm{T}}^{(i)}$ (with $i=2,3,4$) will offer a
   better sensitivity to large \CP phases and may be more suitable for
   experimental analysis.}


\begin{document} 
\section{Introduction}
\label{sec:intro}
The LHC era is just beginning. Flavour physics will play an important
complementary role to direct searches for the theory that lies beyond
the standard model (SM). One central strategy in this period is to
construct observables that are mostly sensitive to specific types of
new physics (NP), in such a way that a deviation could immediately
provide information on the type of NP required: isospin breaking NP,
presence of right-handed currents, scalars, etc. It is essential to
work in a bottom up approach in the direction of constructing a
decision tree that help us to discern which features the NP
model must incorporate and then try to match them into a group of
models.

Few decays are able to provide such a wealth of information with
different observables as \BdbKsll, ranging from forward-backward
asymmetries~(\AFB) and isospin asymmetries to a large number of
angular observables. Each of these observables can provide information
on the different types of NP mentioned above. First published results
from \belle~\cite{:2009zv} and \babar~\cite{Aubert:2006vb} based on
$\order(100)$ decays already demonstrate their feasibility.

In the early years of LHC running one will be restricted to those
observables that may be extracted from the angular distribution using
relatively simple analyses. A study of those observables relevant for
the first few \invfb may be found in \cite{Bharucha:2010bb}. However,
once enough statistics have been accumulated to perform a full angular
analysis based on  the  full 4-body decay distribution of the \BdbKsll,
one has the freedom to design observables with reduced theoretical
uncertainties and specific NP sensitivity.

In~\cite{Kruger:2005ep}, it was proposed  to  construct observables that
maximise the sensitivity to contributions driven by the
electro-magnetic dipole operator \Opep7, while, at the same time,
minimising the dependence on the poorly  known  soft form factors. This
led to the construction of the observable \AT2, based on the parallel
and perpendicular spin amplitudes of the \Kstarz.
The basic idea behind the construction of the observable was inspired
by the zero point of \AFB when calculated as a function of the
dilepton mass squared, \qsq. The zero point has attracted a lot of
attention because of its cleanliness; only at that point one gets a
complete cancellation at LO of the form factor dependence and its
precise position may provide information on the fundamental theory
that lies beyond the SM. For \AT2 the soft form factor dependence
cancels at LO, not only at one point, but in the full \qsq region thus
providing much more experimental information. Moreover, the angular
observable is highly sensitive to new right-handed currents driven by
the  operator \Opep7~\cite{Lunghi:2006hc},  to which \AFB
is blind.

Looking for the complete set of angular observables sensitive to
right-handed currents, one is guided to the construction of the
so-called \AT3 and \AT4 which include longitudinal spin
amplitudes~\cite{Egede:2008uy}. The observables $A_{\mathrm{T}}^{(i)}$ (with
$i=2,3,4$) use the \Kstarz spin amplitudes as the fundamental building
block. This provides more freedom to disentangle the information on
specific Wilson coefficients than just restricting oneself to use the
coefficients of the angular distribution as it was recently done in
\cite{Altmannshofer:2008dz}. For instance, \AT2, being directly
proportional to \Cp7 enhances its sensitivity to the type of NP
entering this coefficient.  Moreover, using each coefficient of the
angular distribution instead of selected ratios of them induces a
larger sensitivity to the soft form factors.

The spin amplitudes are not directly observable quantities; to ensure
that a quantity constructed out of the spin amplitudes can be
observed, it is necessary that it fulfils the same symmetries as the
angular distribution. This observation has the
important consequence~\cite{Egede:2008uy} that \AT1 (first proposed
in~\cite{Melikhov:1998cd}) cannot be extracted from the angular
distribution because it does not respect all its symmetries. Only a
measurement of definite helicity distributions would allow it, but
that is beyond any particle physics experiment that can currently be
imagined~\cite{Egede:2008uy}.

To identify all the symmetries of the angular distribution is one of
the main results of this paper. We discuss the counting of all the
symmetries of the distribution in different scenarios, with and
without scalars and with and without mass terms. We explain the
general method of infinitesimal transformations that allow us
to identify all the symmetries, and we develop here in full detail the
explicit form of the four symmetries in the massless case with no
scalars. As an important cross check of this result, we solve
explicitly the set of spin amplitudes in terms of the coefficients of
the distribution, making use of three out of the four symmetries. Two
important consequences of this analysis are: in solving the system one
naturally encounters an extra freedom to fix one of the variables, and
there is a non-trivial constraint between the coefficients of the
angular distribution considered before as free parameters. It is
remarkable that this unexpected constraint is valid for any decay that
has this same structure.

Finally, we provide an illustrative example of the use of the method
of designing observables with an observable called \AT5 that mixes
simultaneously left/right and perpendicular/parallel spin amplitudes
in a specific way that none of the coefficients of the angular
distribution exhibits, opening different sensitivities to Wilson
coefficients.

In the second part of the paper we present a phenomenological analysis
of the various angular observables based on a  QCD
factorisation~(QCDf)  calculation to NLO precision. Recently, a
very detailed analysis of angular quantities of the decay \BdbKsmm in
various NP scenarios~\cite{Altmannshofer:2008dz} and also an analysis
of the NP sensitivities of angular \CP
asymmetries~\cite{Bobeth:2008ij} were presented. In contrast to 
the former work~\cite{Altmannshofer:2008dz}, we do not assume
that the main part of the $\Lambda_{\rm QCD}/m_b$ corrections are inside the QCD
form factors, but use the soft form factors and develop a new ensemble
method for treating these unknown corrections in a systematic way.
The main differences to the latter analysis of \CP violating
observables is the redefinition of the \CP asymmetries in order to
eliminate the  soft form  factor dependence at LO and the
inclusion of the $\Lambda_{\rm QCD}/m_b$ corrections into the error budget,
which turn out to be significant in the presence of new weak phases.

In~\cite{Egede:2008uy} the experimental preparations for an
indirect NP search using these angular observables were worked out,
showing that a full angular analysis of the decay \BdbKsmm at the
\lhcb\ experiment offers great opportunities. We re-evaluate this
analysis in light of the fourth symmetry for the angular distribution
and conclude that it has no effect on the estimated experimental
errors as all observables are indeed invariant under this symmetry. We
extend the experimental sensitivity study to \CP-violating observables
and show that even with an upgraded \lhcb there is no real sensitivity
to \CP-violating NP phases in \C9 and \C{10}.

The paper is organised as follows: Sec.~\ref{sec:framework} briefly
recall the differential distribution in \BdbKsll and the theoretical
framework of  QCDf  and soft-collinear effective theory~(SCET),
Sec.~\ref{sec:sym-and-obs} extends and completes our previous
discussion about symmetries in the angular distribution, its
experimental consequences are discussed in Sec.~\ref{sec:sense}, and
we perform a phenomenological analysis of the \CP-violating and
\CP-conserving observables in Secs.~\ref{sec:CPviolating}
and~\ref{sec:CPconserving} respectively.

\section{Theoretical framework}
\label{sec:framework}

The separation of NP effects and hadronic uncertainties is
the key issue when using flavour observables in a NP search.
Our analysis is based on  QCDf and SCET  and critically examines the NP reach of
those observables via a detailed error analysis including the impact
of the unknown $\Lambda_{\rm QCD}/m_b$ corrections. In order to make the paper
self contained, we briefly recall the various theoretical ingredients
of our analysis.

\subsection{Differential decay distribution}
\label{sec:distribution}
The decay \BdbKsll, with $\Kstarzb \to \Km \pip$ on the mass shell, is
completely described by four independent kinematic variables, the
lepton-pair invariant mass squared, \qsq, and the three angles
$\theta_l$, $\theta_{K}$, $\phi$. Summing over the spins of the final
state particles, the differential decay distribution of\, \BdbKsll can
be written as
\begin{equation}
\label{eq:differential decay rate}
  \frac{d^4\Gamma}{dq^2\, d\cos\theta_l\, d\cos\theta_{K}\, d\phi} =
   \frac{9}{32\pi} J(q^2, \theta_l, \theta_{K}, \phi)\,,
\end{equation}
The dependence on the three angles can be made more explicit:
\begin{eqnarray} 
  &J(q^2, \theta_l, \theta_K, \phi) =\nonumber\\
    &=  J_{1s} \sin^2\theta_K + J_{1c} \cos^2\theta_K
      + (J_{2s} \sin^2\theta_K + J_{2c} \cos^2\theta_K) \cos 2\theta_l + J_3 \sin^2\theta_K \sin^2\theta_l \cos 2\phi 
\nonumber \\       
    & + J_4 \sin 2\theta_K \sin 2\theta_l \cos\phi  + J_5 \sin 2\theta_K \sin\theta_l \cos\phi+ (J_{6s} \sin^2\theta_K +  {J_{6c} \cos^2\theta_K})  \cos\theta_l 
\nonumber \\      
    & + J_7 \sin 2\theta_K \sin\theta_l \sin\phi  + J_8 \sin 2\theta_K \sin 2\theta_l \sin\phi + J_9 \sin^2\theta_K \sin^2\theta_l \sin 2\phi\,.
\end{eqnarray}
As the signs of the expression depend on the exact definition of the
angles, we have made their definition explicit in
Appendix~\ref{App:Kinematics}.

The $J_i$ depend on products of the six complex $K^*$ spin amplitudes,
$\apaLR$, $\apeLR$ and $\azeLR$ in the case of the SM with massless
leptons. Each of these is a function of $q^2$.  The amplitudes are
just linear combinations of the well-known helicity amplitudes
describing the $B\to K\pi$ transition: 
\be\label{hel:trans}
A_{\bot,\|} = (H_{+1}\mp H_{-1})/\sqrt{2}\, , \qquad A_0=H_0 \, .  
\ee 
Two generalisations will be made from the massless case within our
analysis: if the leptons are considered massive the additional
amplitude $A_t$ has to be introduced. And if we allow for scalar
operators, there is a new amplitude $A_S$. Both can be introduced
independently of the other. For the $J_i$ we find the following
expressions (see also \cite{Kruger:1999xa,
  Kim:2000dq,Faessler:2002ut,Kruger:2005ep})~\footnote{The generalizations to the case which includes  scalar operators was recently presented in~\cite{Altmannshofer:2008dz}.}:
\begin{subequations}
  \label{eq:AC-last}
  \begin{align}
    J_{1s} \equiv a & = \frac{(2+\beta_\ell^2)}{4} \left[|\apeL|^2 + |\apaL|^2 + (L\to R) \right]
    + \frac{4 m_\ell^2}{q^2} \re\left(\apeL^{}\apeR^* + \apaL^{}\apaR^*\right),
    \\
    J_{1c} \equiv b & =  |\azeL|^2 +|\azeR|^2  + \frac{4m_\ell^2}{q^2}
    \left[|A_t|^2 + 2\re(\azeL^{}\azeR^*) \right] + \beta_\ell^2 |A_S|^2 ,
    \\
    J_{2s} \equiv c& = \frac{ \beta_\ell^2}{4}\left[ |\apeL|^2+ |\apaL|^2 + (L\to R)\right],
    \\
    J_{2c} \equiv d& = - \beta_\ell^2\left[|\azeL|^2 + (L\to R)\right],
    \\
    J_3 \equiv e& = \frac{1}{2}\beta_\ell^2\left[ |\apeL|^2 - |\apaL|^2  + (L\to R)\right],
    \\
    J_4 \equiv f& = \frac{1}{\sqrt{2}}\beta_\ell^2\left[\re (\azeL^{}\apaL^*) + (L\to R)\right],
    \\
    J_5 \equiv g& = \sqrt{2}\beta_\ell\left[\re(\azeL^{}\apeL^*) - (L\to R)
      - \frac{m_\ell}{\sqrt{q^2}}\, \re(\apaL {A_S^*}+\apaR {A_S^*})
    \right],
    \\
    J_{6s} \equiv h & = 2\beta_\ell\left[\re (\apaL^{}\apeL^*) - (L\to R) \right],
    \\
    J_{6c}  \equiv h^*&  = 4 \beta_\ell  \frac{m_\ell}{\sqrt{q^2}}\, \re \left[ \azeL {A_S^*} + (L\to R) \right],
    \\
    J_7 \equiv j& = \sqrt{2} \beta_\ell \left[\im (\azeL^{}\apaL^*) - (L\to R)
      + \frac{m_\ell}{\sqrt{q^2}}\, {\im}(\apeL {A_S^*}+\apeR {A_S^*})
    \right],
    \\
    J_8 \equiv k& = \frac{1}{\sqrt{2}}\beta_\ell^2\left[\im(\azeL^{}\apeL^*) + (L\to R)\right],
    \\
    J_9 \equiv m& = \beta_\ell^2\left[\im (\apaL^{*}\apeL) + (L\to R)\right] \, ,
  \end{align}
\end{subequations}
with
\begin{equation}
  \label{eq:betaDef}
  \beta_\ell = \sqrt{1 - \frac{4 m_\ell^2}{\qsq}} \, .
\end{equation}
The notations with the letters $a$-$m$ has been included to make the
comparison to~\cite{Egede:2008uy} easier. Note that $J_{6c} = 0$ in
the massless case.

The amplitudes themselves can be parametrised in terms of the seven
$B\to K^*$ form factors by means of a narrow-width approximation. They
also depend on the short-distance Wilson coefficients $\mathcal{C}_i$
corresponding to the various operators of the effective electroweak
Hamiltonian.  The precise definitions of the form factors and of the
effective operators are given in~\cite{Egede:2008uy}.  Assuming
only the three most important SM operators for this decay mode, namely
\Ope7, \Ope9, and \Ope{10}, and the chirally flipped ones, being
numerically relevant, we have~\footnote{ Following common convention, we use the effective Wilson coefficients of these  operators  which include contributions from four-quark operators as well.}
\begin{subequations}
  \begin{align}
 \apeLR &=  N \sqrt{2} \lambda^{1/2} 
 \bigg[ 
 \left\{ 
   (\Ceff9 + \Cpeff9) \mp (\Ceff{10} + \Cpeff{10})
 \right\} \frac{ V(q^2) }{ m_B + m_\kstar} +  \nonumber \\
 &\quad +\frac{2m_b}{q^2} (\Ceff7 + \Cpeff7) T_1(q^2)
 \bigg] \,,\\
\apaLR &= - N \sqrt{2}(m_B^2 - m_\kstar^2) 
            \bigg[ \left\{ (\Ceff9 - \Cpeff9) \mp (\Ceff{10} - \Cpeff{10}) 
            \right\} \frac{A_1(q^2)}{m_B-m_\kstar}  +    \nonumber  \\
& \quad +\frac{2 m_b}{q^2} (\Ceff7 - \Cpeff7) T_2(q^2) \bigg]  \,, \\
\azeLR &= - \frac{N}{2 m_\kstar \sqrt{q^2}}  \,\,  \bigg[ \left\{ (\Ceff9 - \Cpeff9) \mp (\Ceff{10} - \Cpeff{10}) \right\} \times  \nonumber \\   
& \quad \times \left\{ (m_B^2 - m_\kstar^2 - q^2) ( m_B + m_\kstar) A_1(q^2) 
 -\frac{\lambda A_2(q^2)}{m_B + m_\kstar}
\right\}  + \nonumber \\
& \quad  + {2 m_b}(\Ceff7 - \Cpeff7) \left\{
 (m_B^2 + 3 m_\kstar^2 - q^2) T_2(q^2)
-\frac{\lambda}{m_B^2 - m_\kstar^2} T_3(q^2) \right\}
\bigg] \,, \\
A_t &=  N   \lambda^{1/2}   / \sqrt{q^2}   \left\{ 2 (\Ceff{10} - \Cpeff{10}) \right\} A_0(q^2) \,,
\end{align}
\label{SCETKspin}
\end{subequations}
where the $\mathcal{C}_i$ denote the corresponding Wilson coefficients and 
\begin{equation}
  \label{eq:Lambdadef}
  \lambda= m_B^4  + m_{K^*}^4 + q^4 - 2 (m_B^2 m_{K^*}^2+ m_{K^*}^2 \qsq  + m_B^2 \qsq),
\end{equation}
\begin{equation}
N=\sqrt{\frac{G_F^2 \a^2}{3\cdot 2^{10}\pi^5 m_B^3}
|V_{tb}^{}V_{ts}^{\ast}|^2 \qsq \lambda^{1/2}
\sqrt{1-\frac{4 m_\ell^2}{\qsq}}}.
\end{equation}
Finally we note that, if one additionally considers scalar operators then $A_t$ is
modified by the new Wilson coefficients and an additional amplitude, $A_S$,
proportional to the form factor $A_0(q^2)$, is introduced.

\subsection{QCDf/SCET framework}
The \emph{up-to-date} predictions of exclusive modes are based on QCDf
and its quantum field theoretical formulation, soft-collinear
effective theory~(SCET) \cite{Beneke:2001at,Beneke:2004dp}.  The
crucial theoretical observation is that in the limit where the initial
hadron is heavy and the final meson has a large energy
\cite{Charles:1998dr} the hadronic form factors can be expanded in the
small ratios $\Lambda_{\mathrm{QCD}}/m_b$ and
$\Lambda_{\mathrm{QCD}}/E$, where $E$ is the energy of the meson that
picks up the \s quark from the \Bd decay.  Neglecting corrections of
order $1/m_b$ and $\alpha_s$, the seven a-priori independent $B\to
K^*$ form factors reduce to two universal form factors $\xi_{\bot}$
and $\xi_{\|}$ \cite{Charles:1998dr,Beneke:2000wa}.  These relations
can be strictly derived within the QCDf and SCET approach and lead to
simple factorisation formula for the $B \rightarrow K^*$ form factors
\begin{equation}
F_i(\qsq) \equiv H_i \,\xi + \Phi_B\otimes T_i\otimes \Phi_{K^*}
+\order(\Lambda_{\rm QCD}/m_b) \, .
\end{equation}
There is also a similar factorisation formula for the decay
amplitudes. The rationale of such formulae is that the hard vertex
renormalisations ($H_i$) and the hard scattering kernels ($T_i$) are
quantities that can be computed perturbatively so they can be
separated from the non-perturbative functions that go with them; i.e.\
the light-cone wave functions ($\Phi_i$) which are process-independent
and the soft form factors ($\xi$) which enter in several different $B
\to K^*$ processes.

In general we have no means to calculate $\Lambda_{\rm QCD}/m_b$ corrections to
the QCDf amplitudes so they are treated as unknown corrections, with
the method used for this described in the following section. This, in general, 
leads to a large uncertainty of theoretical predictions based on the
QCDf/SCET which we will explore systematically and make manifest in
our phenomenological analysis.

We do not follow here the approach of~\cite{Altmannshofer:2008dz}
where the full QCD form factors are used in the  QCDf 
formulae. There it is {\it assumed} that the main part of the
$\Lambda_{\rm QCD}/m_b$ corrections are inside the QCD form factors, and
additional $\Lambda_{\rm QCD}/m_b$ corrections are just neglected.
Clearly some of 
the $\Lambda_{\rm QCD}/m_b$ corrections could be moved into  the full QCD form factors. However, 
 there is no robust quantitative estimate of the additional  corrections and, thus, 
 it is not allowed to neglect those unknown corrections, especially 
in view of the expected smallness of new physics effects. 

We follow here   
another strategy. We  construct observables in which the soft form factor dependence cancels out at leading order. Then the influence of the soft  form factors to the physics is almost eliminated from the phenomenological  analysis in a controlled way.  On the other hand we make the uncertainty 
due to $\Lambda_{\rm QCD}/m_b$ corrections manifest in our analysis. It is not expected that there are as large as 
$20-30\%$ as in the $B \to \pi\pi$ decay as argued below. 
The inclusion of the $5-10\%$  errors due to the $\Lambda_{\rm QCD}/m_b$ corrections in our analysis is exploratory of its impact on our observables,  even at the risk to be too conservative. 
Obviously, it is this issue which calls for improvement in view of the new physics reach of these modes.

The theoretical simplifications of the QCDf/SCET approach are
restricted to the kinematic region in which the energy of the $K^*$ is
of the order of the heavy quark mass, i.e.~$\qsq \ll m_B^2$. Moreover,
the influences of very light resonances below 1\gevgev question the
 QCDf  results in that region. In addition, the
longitudinal amplitude in the QCDf/SCET approach generates a
logarithmic divergence in the limit $q^2 \rightarrow 0$ indicating
problems in the theoretical description below 1\gevgev
\cite{Beneke:2001at}.  Thus, we will confine our analysis of all
observables to the dilepton mass in the range, $1\gevgev \leqslant q^2
\leqslant 6\gevgev$.
  
Using the discussed simplifications, the $K^*$ spin amplitudes at
leading order in $1/m_b$ and $\alpha_s$ have the very simple forms:
\begin{subequations}
  \begin{align}
\apeLR &=\sqrt{2} N m_B(1- \sh)\bigg[  
(\Ceff9 + \Cpeff9) \mp (\C{10} + \Cp{10})
+\frac{2\hat{m}_b}{\sh} (\Ceff7 + \Cpeff7) 
\bigg]\xi_{\bot}(E_\kstar),   \\
\apaLR &= -\sqrt{2} N m_B (1-\sh)\bigg[
 (\Ceff9 - \Cpeff9) \mp (\C{10} - \Cp{10}) 
+\frac{2\hat{m}_b}{\sh}(\Ceff7 - \Cpeff7) \bigg] \xi_{\bot}(E_\kstar)\, , \\
\azeLR  &= -\frac{Nm_B }{2 \hat{m}_\kstar \sqrt{\sh}} (1-\sh)^2\bigg[ (\Ceff9 - \Cpeff9)  \mp (\C{10} - \Cp{10}) 
+ 2
\hat{m}_b (\Ceff7 - \Cpeff7) \bigg]\xi_{\|}(E_\kstar)\, ,\\
A_t  &= \frac{Nm_B }{ \hat{m}_\kstar \sqrt{\sh}} (1-\sh)^2\bigg[ \C{10} - \Cp{10} \bigg] \xi_{\|}(E_\kstar) \,,
  \end{align}
\end{subequations}
with $\sh = \qsq/m_B^2$, $\hat{m}_i = m_i/m_B$. Here we neglected
terms of $O(\hat{m}_{K^*}^2)$.  The  scalar spin amplitude  $A_S$
is also proportional to $\xi_{\|}(E_\kstar)$ in this limit.

The symmetry breaking corrections of order $\alpha_s$ can be
calculated in the QCDf/SCET approach.  Those NLO corrections are
included in our numerical analysis
following~\cite{Beneke:2001at,Beneke:2004dp}.  They are presented in
the Appendix of~\cite{Egede:2008uy}.

\subsection{Estimating $\boldmath{\Lambda_{\rm QCD}/m_{b}}$ corrections}
\label{sec:LambdaOverMb}
Our observables have reduced theoretical uncertainties due to the
cancellation of the soft form factors. However, the relations used to
make these cancellations are only valid at LO in the $\Lambda_{\rm QCD}/m_{b}$
expansion, and corrections to higher orders are unknown. For these
theoretically clean observables to be useful, the impact of these
corrections on the observables must be robustly bounded. If NP is to
be discovered in \BdbKsll, it must be possible to demonstrate that any
effect seen is indeed NP and not just the effect of an unknown SM
correction.

To evaluate the effect of the $\Lambda_{\rm QCD}/m_{b}$ corrections, we
parametrise each of the \Kstarz spin-amplitudes with some unknown
linear correction,
\begin{equation}
A'_{i} =  A_{i}(1 + C_{i}e^{i\theta_{i}}),
\end{equation} 
where $C_{i}$ is the relative amplitude and $\theta_{i}$ the relative
strong phase. If we vary $C_{i}$ and $\theta_{i}$ within their allowed
ranges, an estimate for the theoretical uncertainty due to these
unknown parameters can be found. In order to make this parametrisation
generic, however, extra terms must be introduced. In principle the
effective Hamiltonian which controls the decay has three terms,
\begin{equation}
\mathcal{H}_{\mathrm{eff}} = \mathcal{H}^{(u)\mathrm{SM}}_{\mathrm{eff}} + \mathcal{H}^{(t)\mathrm{SM}}_{\mathrm{eff}} + \mathcal{H}^{(t)\mathrm{NP}}_{\mathrm{eff}}.
\end{equation}
The first term is very small as it is suppressed by the factor
$\lambda_u =
{V^{\phantom{*}}_{\mathrm{ub}}V^{*}_{\mathrm{us}}}/{V^{\phantom{*}}_{\mathrm{tb}}V^{*}_{\mathrm{ts}}}$
but is responsible for all the SM \CP-violation in the decay; the
second term is responsible for the decay in the SM; and the third adds
possible NP contributions. A fourth possible term
$\mathcal{H}^{(u)\mathrm{NP}}_{\mathrm{eff}}$ generically does not
contribute to the model independent amplitudes and is neglected. Each
of these contributions is generated by different sets of diagrams and
may have different values of $C_{i}$ and $\theta_{i}$. 

Each amplitude
must be modified to include the three sub-amplitudes with their corrections:
\begin{align}
  A' = &  \left[\left(A_{\mathrm{SM}}(\lambda_{u}\neq 0)-A_{\mathrm{SM}}(\lambda_{u} = 0)\right)\times (1 + C_{1}e^{i\theta_{1}})\right] + \nonumber\\
  & \left[A_{\mathrm{SM}}(\lambda_{u}= 0)\times (1 + C_{2}e^{i\theta_{2}})\right] + \nonumber\\
  & \left[\left(A_{\mathrm{Full}}(\lambda_{u}\neq
      0)-A_{\mathrm{SM}}(\lambda_{u} \neq 0)\right)\times (1 +
    C_{3}e^{i\theta_{3}})\right].\label{eqn:subamps}
\end{align}
It is assumed that only a single NP operator is active so as not to introduce extra terms.
In this formalism, the SM \CP-violating, SM \CP-conserving, and NP parts of the amplitude are then allowed
to have independent $\Lambda_{\rm QCD}/m_{b}$ corrections and strong phases.

An estimate of the theoretical uncertainty arising from the unknown
$\Lambda_{\rm QCD}/m_{b}$ corrections and strong phases can now be made using a
randomly selected ensemble. For each member of the ensemble, values of
$C_{1-3}$ and $\theta_{1-3}$ are chosen in the ranges $C_{i} \in
[-0.1,0.1]$ or $C_{i} \in [-0.05,0.05]$ and $\theta_{i} \in
[-\pi,\pi]$ from a random uniform distribution. This is done for the
 seven amplitudes, $A_t$, $A_{0}^{L,R}$,
$A_{\parallel}^{L,R}$, $A_{\perp}^{L,R}$, to provide a complete
description of the decay. It is assumed that the corrections and
phases are not functions of \qsq, although in practise they may
actually be. Any unknown correlations are also ignored. While these
effects could lead to an underestimate of the theoretical envelope, it
is thought that this method allows for a conservative estimate of the
theoretical uncertainties to be made.

To estimate the contribution to the theoretical uncertainties from
$\Lambda_{\rm QCD}/m_{b}$ corrections for a particular observable, each element
in the ensemble was used to calculate the value of that observable at
a fixed value of \qsq. A one $\sigma$ error is evaluated as the
interval that contains 66\% of the values around the median.
This is
done for both $C_{i} \in [-0.05,0.05]$ and $C_{i} \in [-0.1,0.1]$ to
illustrate the effects of five and ten percent corrections. By
repeating this process for different values of \qsq, bands can be
built up.  No assumption
of Gaussian statistics has been made; the bands illustrate the probable range for
the true value of each observable, given the current central value. The method
allows for the probability that a given experimental result is due to an
unknown SM correction to be found.

The choice $|C_i| < 10\%$ is based on a simple dimensional
estimate. We emphasize here that there is no strict argument available
to bound the $\Lambda_{\rm QCD}/m_\b$ corrections this way. But we can state that
the chiral enhancement of $\Lambda_{\rm QCD}/m_\b$ corrections in the case of
hadronic \B decays does not happen in the case of the semileptonic
decay mode with a {\it vector} final state.

The process described here avoids any assumptions about correlations
between the corrections and is thus statistically more rigorous than
what was done in~\cite{Egede:2008uy}, where corrections to amplitudes
were considered one by one and then added in quadrature. 
The $\Lambda_{\rm QCD}/m_{b}$ bands it produces
are reduced when compared to those of~\cite{Egede:2008uy}. 
It also allows us to investigate the effect of the $\Lambda_{\rm QCD}/m_{b}$ corrections
for \CP-violating observables.

\section{Symmetries and observables}\label{sec:sym-and-obs}

The experimental degrees of freedom determined by the $J_i$ terms and
the theoretical degrees of freedom determined by the spin amplitudes $A_j$
have to match. There are two effects to consider for this: different
values of the $A_j$ can give rise to the same differential
distribution Eq.~(\ref{eq:differential decay rate}) and thus cannot be
distinguished; and in some cases the experimental coefficients are not
independent, meaning that not all arbitrary values of the $J_i$ are
possible. The first effect we call a continuous symmetry
transformation. For the degrees of freedom to match we
have
\begin{equation}
  \label{eq:nDOFmatch}
  n_c - n_d = 2n_A - n_s \,,
\end{equation}
where $n_c$ is the number of coefficients in the differential
distribution (the number of $J_i$), $n_d$ the number of dependencies
between the different coefficients, $n_A$ the number of spin
amplitudes (the  $A_j$, each is complex and hence has two degrees of
freedom), and $n_s$ the number of continuous symmetries.

We considered this situation in our previous
paper~\cite{Egede:2008uy} for the case of massless leptons and return
to it again here. It is easy to see that in the massless limit, $J_{1s}
= 3 J_{2s}$ and  $J_{1c} = - J_{2c}$. What is not so obvious is that $J_9$
can be expressed in terms of the other 8 remaining coefficients. Going
back to Eq.~(\ref{eq:nDOFmatch}) it can be seen that the massless case
in fact must have 4 symmetries and not 3 as we claimed in the previous
paper.

Below we first outline how the symmetries and dependencies can be
identified before we move onto their explicit form and the
interpretation.

\subsection{Infinitesimal symmetries}
\label{sec:DifferentialSymmetries}
By an infinitesimal symmetry is meant one where the theoretical spin
amplitudes $A_j$ are changed in an infinitesimal way leaving the
$J_i$ coefficients in Eq.~(\ref{eq:AC-last}) unchanged. The
infinitesimal symmetries will define a system of coupled ordinary
differential equations that, if solved, are the global symmetries we
look for. There is no guarantee that these symmetries will allow for
the continuous transformation between two arbitrary  sets  of amplitudes
which have the identical angular distribution; there could in
principle be several disjoint regions separated by divergences.

If we, in this example, look at massless leptons and ignore the scalar
amplitude, the coefficients of the spin amplitudes can be defined as a
vector $\vec{A}$ with 12 components 
\begin{eqnarray}
  \label{eq:Avector}
  \vec{A} & = & 
      \left(\Re(\apeL),\Im(\apeL),\Re(\apaL),\Im(\apaL),\Re(\azeL),\Im(\azeL), 
      \phantom{)}\right. \nonumber \\ 
      & &\left. \phantom{(}  \Re(\apeR),\Im(\apeR),\Re(\apaR),\Im(\apaR),\Re(\azeR),\Im(\azeR)\right)
\end{eqnarray} 
corresponding to the real and imaginary parts of the amplitudes. For
each of the coefficients $J_i$ we can find the derivative with respect
to the spin amplitudes. As an example 
\begin{equation}
  \label{eq:CoefficientGradient}
  \vec{\nabla}(J_{1c}) =
  \left(0,0,0,0,2 \Re(\azeL), 2 \Im(\azeL),0,0,0,0,2 \Re(\azeR),2 \Im(\azeR)\right) \, .
\end{equation} 
There will be eleven such gradient vectors in the massless case, as
$J_{6c}=0$.

Now, any infinitesimal transformation can be written on the form
\begin{equation}
  \label{eq:InfinitesimalTransformationDef}
  \vec{A'} = \vec{A} + \delta \vec{s} \, .
\end{equation}
For the infinitesimal transformation to leave the coefficients
unchanged, the vector $\delta \vec{s}$ has to be perpendicular to the
hyperplane spanned by the set of gradient vectors. Or in other words,
$\delta \vec{s}$ represents a symmetry \emph{if, and only if}
\begin{equation}
  \label{eq:InfinitesimalSymmetryDef}
  \forall i \in J_i : \vec{\nabla}_i \perp \delta \vec{s} \, .
\end{equation}

Looking back at Eq.~(\ref{eq:nDOFmatch}) we have, for the massless case,
$n_c=11$. If the $J_i$ were all independent the gradient vectors would
span an 11 dimensional hyperplane. In fact, it turns out that they only
span 8 dimensions\footnote{Any program able to handle symbolic algebra
  will be able to show this.}, which shows that there are three
dependencies between the $J_i$'s, giving $n_d=3$. As we have $n_A=6$
from the amplitudes we see from Eq.~(\ref{eq:nDOFmatch}) that we have
$n_s=4$ corresponding to 4 symmetries. For the dependencies, only the
first two $J_{1s} = 3 J_{2s}$ and  $J_{1c} = - J_{2c}$  are trivial; the
third one we derive in the next section.

\subsection{Explicit form of symmetries}
\label{sec:ExplicitSymmetry}
It is helpful for the discussion to make the following definitions.
\begin{subequations}
  \begin{align}
  n_1&=(\apaL, \apaR^*) \,,\\
  n_2&=(\apeL, - \apeR^*) \,, \\
  n_3&=(\azeL, \azeR^*) \,,
  \end{align}
\end{subequations}
or in terms of helicity amplitudes
\begin{subequations}
  \begin{align}
  m_1&=\frac{1}{\sqrt{2}} (n_1+n_2)=(H_{+1}^{L},{H_{-1}^{R}}^* ) \,, \\
  m_2&=\frac{1}{\sqrt{2}} (n_1-n_2)=(H_{-1}^{L},{H_{+1}^{R}}^*) \,, \\
  m_3&=n_3=(H_{0}^{L},{H_{0}^{R}}^*) \, .
  \end{align}
\end{subequations}
In fact, all the information of the angular distribution is
encoded in the moduli of the three $n_i$ vectors and their relative
complex scalar products:
\begin{eqnarray} \label{eq1} |n_1|^2=\frac{2}{3} J_{1s} - J_3 \,, \quad
  \quad |n_2|^2&=&\frac{2}{3} J_{1s} + J_3 \,, \quad \quad 
  |n_3|^2=J_{1c} \,, \\ 
  n_1 \cdot n_2=\frac{J_{6s}}{2}- i J_9 \,, \quad \quad 
  n_1 \cdot n_3&=&\sqrt{2} J_4 - i \frac{J_7}{\sqrt{2}} \,, \quad \quad
  \label{eq6}
  n_2 \cdot n_3=\frac{J_5}{\sqrt{2}} - i \sqrt{2} J_8 \,,
\end{eqnarray}
where $n_i$ being a complex vector implies that the scalar product is
$n_i\cdot n_j= \sum_k n_{i_k} n_{j_k}^* $. The coefficients $J_{2s}$
and $J_{2c}$ are absent because they are obviously redundant.

The differential distribution is invariant under the following
four independent symmetry transformations of the amplitudes
\begin{equation}
  \label{eq:SymMassless}
  n_i^{'} = 
  \left[
    \begin{array}{ll}
      e^{i\phi_L} & 0 \\
      0 & e^{-i \phi_R}
    \end{array}
  \right]
  \left[
    \begin{array}{rr}
      \cos \theta & -\sin \theta \\
      \sin \theta & \cos \theta
    \end{array}
  \right]
  \left[
    \begin{array}{rr}
      \cosh i \tilde{\theta} &  -\sinh i \tilde{\theta} \\
      -\sinh i \tilde{\theta} & \cosh i \tilde{\theta}
    \end{array}
  \right]
  n_i \,,
\end{equation}
where $\phi_L$, $\phi_R$, $\theta$ and $\tilde{\theta}$ can be varied
independently.  Identical  transformations can be carried out on the
$m_i$. Normally, there is the freedom to pick a single global phase,
but as $L$ and $R$ amplitudes do not interfere here, two phases can be
chosen arbitrarily as reflected in the  first transformation matrix.

The interpretation of the third and fourth symmetry is that they
transform a helicity $+1$ final state with a left handed current
into a helicity $-1$ state with a right handed current. As we
experimentally  cannot  measure the simultaneous change of helicity and
handedness of the current, these transformations turn into
symmetries for the differential decay rate.

\subsection{Relationship between coefficients in 
  differential distribution}
\label{sec:Dependency}
As was mentioned earlier, we have identified an extra dependency among
the coefficients in the massless case. Here we outline how it can be
derived. 

If we use the two global phase symmetry transformations we can rotate
the vector $n_1$ to make it real (\apaL\ and \apaR\ become
real).\footnote{Indeed the system can also be solved using only one of
  the two global symmetries and keep \apaR complex.} We can
then choose the angle $\theta$ of the third symmetry to make
$\apaL=0$. Notice that we have not made use of the fourth
symmetry. The implications of this fourth symmetry will become
manifest when solving the system. With these choices
\begin{equation}
  n_1=(0,\apaR) \,,
\end{equation}
where \apaR is a positive real parameter. Using three
of Eqs.~(\ref{eq1})-(\ref{eq6}) together with the symmetries, one can
determine four of the spin amplitudes (their moduli and phases):
\begin{subequations}
  \begin{align}
\apaL &=0   \,, \\
\apaR &=\sqrt{|n_1|^2}=\sqrt{\frac{2}{3} J_{1s} - J_3} \,, \\
\apeR &=- \frac{n_1 . n_2}{\sqrt{|n_1|^2}}= - \frac{\left(J_{6s} - 
    2  i J_9 \right)}{2 \sqrt{\frac{2}{3} J_{1s} - J_3}}  \,, \\
\azeR &=\frac{n_1 . n_3}{\sqrt{|n_1|^2}}=  \frac{2 J_4 - i 
  J_7}{
  \sqrt{\frac{4}{3} J_{1s}-2 J_3}} \, . \label{eq8}
  \end{align}
\end{subequations}
The remaining three equations from Eqs.~(\ref{eq1})-(\ref{eq6})
determine, on one side, the moduli of \apeL and \azeL,
\begin{subequations}
  \begin{align}
    |\apeL|^2 &=|n_2|^2- \frac{|(n_1 . 
n_2)|^2}{|n_1|^2}=\frac{\frac{4}{9} 
J_{1s}^{2}-J_3^2 - \frac{1}{4} J_{6s}^{2} - J_9^2}{\frac{2}{3} J_{1s} - 
J_3} \,, \\
|\azeL|^2 &=|n_3|^2 - \frac{|(n_1 . n_3)|^2}{|n_1|^2}=
\frac{J_{1c} \left(\frac{2}{3} J_{1s} - J_3 \right) - 2 J_4^2- \frac{1}{2} 
J_7^2}{\frac{2}{3} J_{1s} -
J_3} \,,
 \label{eq12}
  \end{align}
\end{subequations}
and on the other, the phase difference corresponding to the previous two 
amplitudes:
\begin{align}\label{phases}
e^{i(\phi_{\perp}^{L}-\phi_{0}^{L})} &= \frac{ (n_2 \cdot n_3) 
|n_1|^2 
- 
(n_2 \cdot n_1) (n_1 \cdot n_3)}{\left[
\left(|n_1|^2 |n_2|^2-|(n_2 \cdot 
n_1)|^2\right)\left(|n_1|^2|n_3|^2-|(n_3 \cdot
n_1)|^2\right)\right]^{1/2}} \nonumber \\
&=\frac{ J_5 \left(\frac{2}{3} J_{1s}-J_3 \right)-J_4 J_{6s} -J_7 
J_9 
-i \left(\frac{4}{3} J_{1s} J_8 - 2 J_3 J_8 + 2 J_4 J_9 - \frac{1}{2} 
J_{6s} J_7 \right) }{\left[
2 \left(\frac{4}{9} J_{1s}^{2} -J_3^2- \frac{1}{4} J_{6s}^{2}-J_9^2 
\right) \left(J_{1c} \left( \frac{2}{3} J_{1s} - J_3 \right) - 2 J_4^2 - 
\frac{1}{2} J_7^2 \right)\right]^{1/2}} \, .
\end{align}

\enlargethispage{\baselineskip}
Here is where the fourth symmetry becomes manifest. On one side, this 
equation tells us that you have the freedom to choose one of the two 
phases $\phi_{\perp}^L$ or $\phi_{0}^L$ to zero. On the other side, 
given that the LHS of the previous equation is a pure phase, the 
 modulus  of the RHS should be one. This implies the following 
important non-trivial relationship between the coefficients of the 
distribution
\begin{align}
J_{1c} = & \ 6\frac{ (2 J_{1s}+ 3 J_3) \left(4 J_4^2+J_7^2\right)
+ ( 2 J_{1s} - 3
J_3)
\left(J_5^2+4 J_8^2 \right)}{16 J_1^{s \, 2} -
9
\left(4 J_3^2+ J_6^{s \, 2} + 4 J_9^2 \right)} \nonumber \\
& \ -36 \frac{J_{6s} (J_4 J_5 +
 J_7 J_8) + J_9 (J_5 J_7 - 4 J_4 J_8)}{16 J_{1s}^{2} -
9
\left(4 J_3^2+ J_{6s}^{2} + 4 J_9^2 \right)} .
\end{align} 
It is important to remark that this equation can be very easily
generalised to the massless case with scalars using the relations
$J_{1s}=3 J_{2s}$ and $J_{1c}=-J_{2c}$ in the previous equation. Also
the massive case with no scalars can be included by introducing the
$\beta$ factors inside the $J_i$ coefficients. There is no such
equation in the massive case with scalars due the fact that the number
of coefficients of the experimental distribution is identical to the
number of theoretical amplitudes and symmetries (see
Tab.~\ref{tab:symmetries}).

\subsection{Experimental issues}
\label{sec:SymExp}
The symmetries discussed above can be used to fix the spin-amplitude
components by choosing specific values of the relevant rotation
angles. We give an explicit example of this for the case where the
lepton mass is neglected. We choose to make the following constraint:
\begin{equation}
  \Re(\apaL)=\Im(\apaL)=\Im(\apaR)=\Im(\apeL)=0.
\end{equation}

This can be achieved by first performing the last transformation, shown in
\eq{eq:SymMassless}, with the value of $\tilde{\theta}$ given by:
\begin{align}
  \sin \tilde{\theta}=&\sqrt{\frac{z-1}{2z}} \,, \qquad\qquad\quad
  \cos\tilde{\theta}=\sqrt{\frac{z+1}{2z}} \,, \\
  \intertext{where} z= & \sqrt{1 + 4 {\left[\frac{
          \Re(\apaL)\Im(\apaR)+\Re(\apaR)\Im(\apaL)}{\Re(\apaR)^{2}+\Im(\apaR)^{2}- \Re(\apaL)^{2} - \Im(\apaL)^{2}}\right]}^2}\, .
\end{align} 
Next, the third rotation angle, $\theta$,  is used again  in \eq{eq:SymMassless}:
\begin{equation}
  \tan\theta =\frac{\sqrt{1+z}\, \Re(\apaL)-\sqrt{z-1}\, \Im(\apaR)}{\sqrt{1+z}\, \Re(\apaR)+\sqrt{z-1}\, \Im(\apaL)} \, .
\end{equation}
The $L$-fields are phase-shifted by $\phi_L$:
\begin{equation}
  \tan\phi_L= - \frac{\cos\tilde{\theta}[\cos\theta\;\Im(\apeL)-\sin\theta\;\Im(\apeR)]+\sin\tilde{\theta}[\cos\theta\;\Re(\apeR)+\sin\theta\;\Re(\apeL)]}{\sin\tilde{\theta}[\cos \theta\;\Im(\apeR)-\sin\theta\;\Im(\apeL)]+\cos\tilde{\theta}[\cos \theta\;\Re(\apeL)+\sin \theta\;\Re(\apeR)]} \,,
\end{equation}
and finally the last $R$-field transformation can be performed
substituting $(\perp \to \|)$ and $(L \leftrightarrow R)$ into the
previous expression:
\begin{equation}
\tan\phi_R=- \frac{\cos\tilde{\theta}[\cos \theta\;\Im(\apaR)-\sin \theta\;\Im(\apaL)]+\sin\tilde{\theta}[\cos \theta\;\Re(\apaL)+\sin \theta\;\Re(\apaR)]}{\sin\tilde{\theta}[\cos \theta\;\Im(\apaL)-\sin \theta\;\Im(\apaR)]+\cos\tilde{\theta}[\cos \theta\;\Re(\apaR)+\sin \theta\;\Re(\apaL)]} \, .
\end{equation}

\subsection{Constructing observables}
\label{sec:NewObs}
In~\cite{Kruger:2005ep,Egede:2008uy}, as well as here, we use the spin
amplitudes to construct our observables. There are two main advantages
of this approach, one is experimental and the other is theoretical. On
the experimental side, we have found that fitting directly the angular
coefficients $J_i$, without taking into account the relations between
them, leads to fit instabilities. These  relations, coming from the
underlying \Kstarz spin  amplitudes, can be found in
Sec.~\ref{sec:Dependency}. The theoretical argument has to do with our
aim at constructing observables that fulfil certain criteria, namely
maximal sensitivity to a specific NP operator, like new right-handed
currents, and minimal sensitivity to poorly known form factors. Given
that our main tools are the spin amplitudes, it is a
straight-forward exercise to design observables with a specific NP
sensitivity and small hadronic uncertainties. We also have more
freedom to construct observables than just using each coefficient of
the distribution as an observable. As the spin amplitudes can be
extracted directly in the full-angular analysis, there is no penalty
on the final experimental uncertainty from using a non-trivial
functional form to make the observable.

The symmetries of the angular distribution play a crucial role in our
approach.  Once a quantity has been designed, it is a necessary
condition for being an observable based on the angular distribution
that it respects all the symmetries of this distribution. For example
in~\cite{Egede:2008uy}, we have shown explicitly that a
previously discussed transversity amplitude \AT1 does not
fulfil all the symmetries of the angular distribution.  This implies
that this quantity cannot be measured at the LHCb experiment or  at
future  super-$B$ factory experiments; a measurement of the spins of
the final-state particles would be required for that.

Let us finally discuss a new \CP-conserving observable that we call
\AT5. It is defined as: 
\begin{equation}
 \label{eq:AT5general}
 \AT5 = \frac{\big|\apeL \apaR^* + \apeR^* \apaL\big|}
             {\big|\apeL\big|^2 + \big|\apeR\big|^2 + \big|\apaL\big|^2 + \big|\apaR\big|^2}   \, .
\end{equation}
It probes the transverse spin amplitudes $A_{\bot}$ and
$A_{\parallel}$ in a different way than \AT2. Direct inspection of
Eq.~(\ref{eq:AC-last}) shows that there is no single angular
coefficient mixing $L$ with $R$ and $\perp$ with $\parallel$
simultaneously in the way \AT5 does.

It is a simple exercise to check that this observable fulfils the four
symmetries described in Eq.~(\ref{eq:SymMassless}). Once this
invariance is fulfilled\footnote{Notice one could try to write \AT1 in
  terms of the $J_i$ using the explicit solution, but this is not
  allowed since \AT1 is not invariant \cite{Egede:2008uy}.}  one is
allowed to use the explicit solution in the massless case provided in
the previous subsection Eqs. (\ref{eq1})-(\ref{eq6}):
\begin{equation}
\AT5 \Big\vert_{m_{\ell}=0}= \frac{\sqrt{16 J_{1}^{s\,2}-9 J_{6}^{s \,2} - 36 (J_{3}^2+J_{9}^2)}}{8 J_{1}^{s}} \, .
\end{equation}
A discussion on the properties and sensitivities of this observable is
presented in Sec.~\ref{sec:CPconserving}.

\subsection{More general cases}
\label{sec:MoreGeneralSymmetries}
The discussion of the differential symmetries from
Sec.~\ref{sec:DifferentialSymmetries} can be generalised to the cases
where the leptons are no longer considered massless and where a scalar
amplitude is included:
\begin{description}
\item[Massless leptons with scalars] The inclusion of the scalar
  amplitude $A_S$, gives us seven amplitudes. The four explicit
  symmetries in Eq.~(\ref{eq:SymMassless}) are still valid and we have
  in addition
  \begin{equation}
    \label{eq:SymScalar}
    A^{'}_{S} = e^{i \phi_S} A_{S} \,,
  \end{equation}
  expressing that the phase of $A_S$ cannot be determined.
\item[Massive leptons without scalars] We have the seven amplitudes
  $\apeLR$, $\apaLR$, $\azeLR$ and $A_t$ in this case and still eleven
  coefficients. As a fact of elementary quantum mechanics we still
  have a global phase transformation corresponding to $\phi_L=\phi_R$,
  but the other two symmetries from the massless case are no longer
  valid. There is a new symmetry concerning the phase of $A_t$ given
  as:
  \begin{equation}
    \label{eq:SymT}
    A^{'}_{t} = e^{i \phi_t} A_{t} \, .
  \end{equation}
  This leaves us with two symmetries where only the differential form
  is known.
\item[Massive leptons with scalars] We now have all eight amplitudes
  and, with the inclusion of $J_{6c}$, we have twelve coefficients. The
  global phase transformation, $\phi_L=\phi_R$, and the phase
  transformation of $A_t$ in Eq.~(\ref{eq:SymT}) are still valid. In this
  case, there is no dependency between any of the coefficients,  leaving
  us with two symmetries where only the differential form is known.
\end{description}
So while we in some cases only know the differential form of the
symmetries, we are still able to test if observables respect the
symmetries (see Sec~\ref{sec:NewObs}) and we can also determine the
optimal set of amplitudes to fit for in an experimental fit (see
Sec.~\ref{sec:SymExp}). In Tab.~\ref{tab:symmetries} we summarise the
full knowledge about the symmetries.
\TABLE{
  \caption{The dependencies between the coefficients in the differential
    distribution and the symmetries between the amplitudes in several
    special cases.}
  \vspace{5mm}
  \begin{tabular}{ccccc}
    Case & Coefficients & Dependencies & Amplitudes & Symmetries \\
    \hline
    $m_\ell=0$, $A_S=0$ & 11 & 3 & 6 & 4 \\
    $m_\ell=0$        & 11 & 2 & 7 & 5 \\
    $m_\ell>0$, $A_S=0$ & 11 & 1 & 7 & 4 \\
    $m_\ell>0$        & 12 & 0 & 8 & 4
  \end{tabular}
  \label{tab:symmetries}
}

\section{Experimental Sensitivities}\label{sec:sense}

In~\cite{Egede:2008uy}, a fitting technique was investigated that
allowed the extraction of the \Kstarz spin amplitudes from the full
angular distribution in the massless lepton limit. \eq{eq:differential
  decay rate} can be interpreted as a probability density function
(PDF) and normalised numerically. We parametrise it in terms of six
complex \Kstarz spin amplitudes, which are functions of \qsq only. In
the limit of infinite experimental data, and for a fixed value of \qsq,
these amplitudes can be found by fitting the relative contribution of
each angular coefficient as a function of the three decay angles. As
discussed in \sec{sec:sym-and-obs}, the symmetries of the distribution
can then be used to reduce the number of unknowns; if we consider the
real and imaginary amplitude components separately, the twelve
parameters can be reduced to eight using the symmetry constraints. A
further spin-amplitude component may be removed by noting that
\eq{eq:differential decay rate} is only sensitive to relative
normalisations. This leaves seven free parameters at each point in
\qsq. However, in~\cite{Egede:2008uy}, only three, out of four,
symmetry constraints were considered meaning that, in principle, the
fits presented were under-constrained. The implications of this will
be investigated in this section.

Despite the large increases in \BdbKsmm statistics expected at \lhcb,
the number of signal events available will still be too small for a
fixed \qsq approach to be taken. Instead, the spin-amplitude
components are parametrised as second-order polynomials in the region
\(\qsq\in [1,6]\gevgev\). These are normalised relative to the
value of $\Re(\azeL)$ at a fixed value, \(X_{0}\),  of \qsq. Rather than
fitting directly for the amplitudes, we aim to extract the
coefficients of these polynomials. This introduces a number of model
biases: the underlying spin amplitudes are assumed to the smoothly
varying in the \qsq window considered. As noted in~\cite{Egede:2008uy},
this was verified for a number of NP models. There is also an implicit
assumption that the \qsq-dependent shape of the spin amplitudes is
invariant under the symmetries of the angular distribution. Neglecting
background parameters, the \qsq-dependent fit has
\(((12-4)\times 3) - 1=23\) free parameters to
be extracted, or 26 in~\cite{Egede:2008uy}. These will be
labelled the four- and three-symmetry fits respectively.

The three-symmetry fit although, in principle, under-constrained is
able to converge due to the polynomial parametrisation employed. By
requiring that three of the spin amplitude components vanish for all
values of \qsq, we have used our freedom to choose values of
\(\phi_{L}\), \(\phi_{R}\), and \(\theta\) from
Eq.~(\ref{eq:SymMassless}) at each point in \qsq; the value of
\(\tilde{\theta}\) is still free to vary. However, the PDF,
\eq{eq:differential decay rate}, is invariant under changes of
\(\tilde{\theta}\); hence, the negative log-likelihood (NLL) used
during minimisation should not be sensitive to its value. The \qsq
dependent shape of each amplitude component is manifestly not
invariant under changes in \(\tilde{\theta}\) -- the rotation it
implies mixes the imaginary parts of the left- and right-handed
amplitudes. The polynomial parametrisation of the spin-amplitude
components requires that each amplitude must be smoothly varying. The
fit then selects the value of \(\tilde{\theta}\) for each signal event
which produces the most polynomial-like distribution, as this will
have the smallest NLL. The general minimising algorithm employed is
then able to find a genuine minimum and converge properly; the
imposition of the polynomial ansatz allowed the under-constrained fit
of~\cite{Egede:2008uy} to converge properly. As the experimental
observables are invariant under all four symmetries, their \qsq
dependent distributions can be found correctly; there are no
significant biases seen in the central values extracted compared to
the input distribution. Small biases {\em are} seen in the individual
spin-amplitude components; with hindsight, correlations between these
components were induced by the presence of the fourth symmetry.

\subsection{Experimental Analysis}\label{sec:sensitivity}

The discussion above explains why the three-symmetry fit is able to converge
successfully, and suggests that there should be no change in the
experimental uncertainties found when the extra symmetry constraint is added. It is
important to demonstrate that this is the case. As before, the experimental sensitivity
to different observables can be estimated using a toy Monte Carlo (MC) approach and used to 
compare the three- and four-symmetry fits.  

\subsubsection{Generation}\label{sec:generation}
An ensemble of data sets for \BdbKsmm can be generated; each data set
contains the Poisson fluctuated number of signal and background events
expected after \lhcb has collected 10\invfb of integrated
luminosity. Estimates of the signal and background yields were taken
from~\cite{:2009ny,Patel:1157434} and scaled linearly. The signal
distribution was generated using the \Kstarz spin amplitudes discussed
in \sec{sec:framework} as input. The contribution from terms including
the muon mass were included. No assumption of polynomial variation of
the amplitudes was used in the generation. The signal is assumed
to have a Gaussian distribution in $m_\B$ with a width of $14\mev$ in
a window of $m_\B \pm50\mev$ and a Breit-Wigner in $m_{K\pi}$ with
width $48\mev$ in a window of $m_{\Kstarz} \pm 100\mev$. A simplified
background model is included; it is flat in all decay angles,
effectively treating all background as combinatorial, but follows the
\qsq distribution of the signal. Detector acceptance effects as
described in \cite{:2009ny} are not taken into account. When
considering \CP-conserving quantities, the \B and \Bb samples are
simply considered together.  We do not include any contributions from
non-resonant \mbox{$\Bdb \to \Km\pip\mup\mun$}.

\subsubsection{Observable sensitivities}\label{sec:decay_model}
The ensemble of simulated data sets can then be used to estimate the
experimental uncertainties expected for a given integrated luminosity
at \lhcb. For each data set, the full angular fit was performed to
find the most likely value for each of the free parameters for that
data set. For the three-symmetry fit there were 27 free
parameters; 26 for the signal distribution and one to describe
the level of background seen. For the four-symmetry fit, only 24
parameters were required. In total we created an ensemble of 1200
experiments and will, thus, at a given value of \qsq, get 1200
different determinations of each observable. By looking at the point
where 33\% and 47.5\% of results lie within either side of the median
of the results we can form asymmetric $1\sigma$ and $2\sigma$
errors. Connecting these at different \qsq values gives us $1\sigma$
and $2\sigma$ bands for the experimental errors on the observable.

\subsubsection{\CP asymmetries}
The sensitivity to various \CP asymmetries was also considered. In
this case, separate \B and \Bb samples were generated and fit
independently. Each sample had on average half the number of signal
and background events as those described in \sec{sec:generation}. The
results of a \B and a \Bb fit could then be combined by re-normalising
the \B amplitudes found, so that the extracted value of $\Re(\azeL)$ at
\(X_{0}\) was the same in both samples. This gives sensitivity to \CP
asymmetries relative to this point. By considering many \B and \Bb
samples together, estimates of the experimental sensitivity to the \CP
asymmetries could then be found. In a real measurement, a more
sophisticated approach would be taken which considered the two samples
simultaneously; however, our simplified approach gives a
reasonable first estimate of the experimental sensitivities obtainable
and allow comparison with theoretical requirements.

\subsection{The polynomial ansatz re-examined}

A key assumption of the fitting approach taken in~\cite{Egede:2008uy}
is that the spin-amplitude components are smoothly varying functions
in the range $\qsq \in [1,6]\gevgev$. It was found that when all four
symmetries of the massless angular distribution are taken into
account, this assumption {\em no longer holds}; indeed the shape of
the spin-amplitude components is not invariant under the four
symmetries and their shape can be distorted so they are no longer well
described by second-order polynomials. Other parameterization choices
are likely to be equally vulnerable to these problems unless
they are explicitly invariant under all symmetries of the angular distribution.
Consider the three-symmetry
case at a fixed \qsq value: in~\cite{Egede:2008uy}, $\azeR$ is removed
by setting $\theta = \arctan(-\azeR/\azeL)$ once their phases have
been rotated away. This can be understood by substituting the
trigonometric identities,
\begin{equation}
\sin(\arctan(\theta)) = \frac{\theta}{\sqrt{1 + \theta^{2}}} \,, \quad \cos(\arctan(\theta)) = \frac{1}{\sqrt{1 + \theta^{2}}} \, ,
\end{equation}
into Eq.~(\ref{eq:SymMassless}). This introduces a $[1 +
({\azeR}^{2}/{\azeL}^{2})]^{-\frac{1}{2}}$ term into each non-zero
amplitude component, which will not be well behaved as $\azeL \to
0$. For the three-symmetry fit, these problems can be avoided by
taking $\Re(\azeL)$ as the reference amplitude component, forcing it to
be relatively large at $X_{0}$. However, to include the fourth
symmetry constraint, a more complicated form must be used in order to
set four amplitude components simultaneously. A different value of
each of the four rotation angles is required for every point in \qsq
due to the changing spin amplitudes. There is no guarantee that a set
of rotation angles can be found such that the unfixed spin-amplitude
components resemble smoothly varying polynomials for all \qsq. The
\qsq dependence of the SM input amplitude $\Re(\azeL)$ is shown in
\fig{fig:imA0L-T} once the four symmetries have been applied to fix
$\Im(\apaL)$, $\Im(\apaR)$, $\Re(\apaL)$, and $\Im(\apeL)$ to zero, as
required for in the next section. This particular feature is caused by
$\Re(\apaL)\to 0$ at $\qsq \approx 2\gevgev$; other rotation choices lead to
similar features. The distribution can no longer be well described by
a second-order polynomial. It may be possible to find a choice of
rotation parameters that preserve the polynomial features of the input
spin-amplitude components, however, there are no guarantee that a
particular choice would work when faced with experimental
data. Indeed, an incorrect choice will lead to biases in the case
where the parametrisation is a poor match for the underlying
amplitudes. A more generic solution is required and could form the
basis for further investigations.

\EPSFIGURE{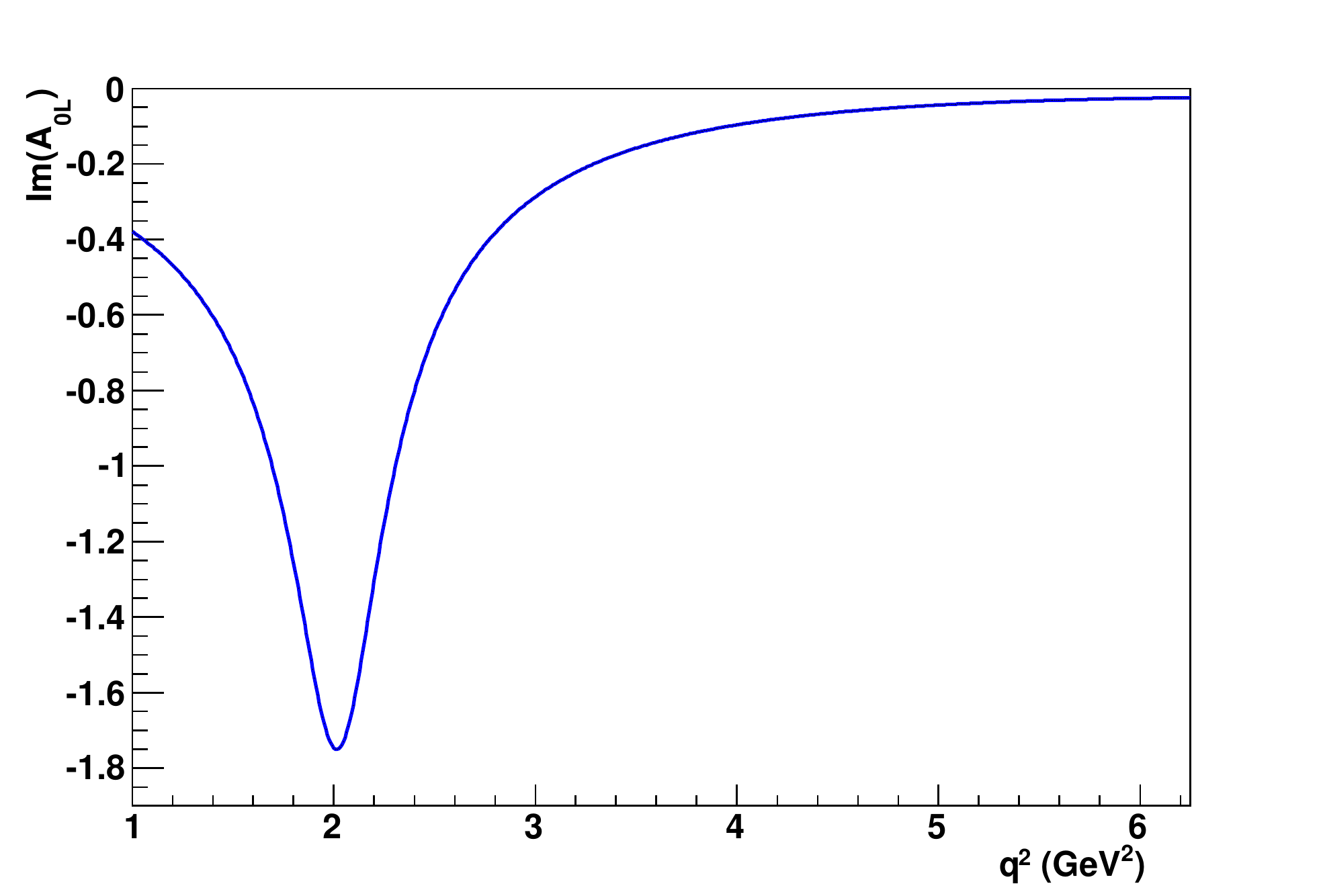,width=0.48\textwidth}{\label{fig:imA0L-T}The \qsq
  dependence of $\Im(\azeL)$ after using the four symmetries of the
  full-angular distribution to fix $\Im(\apaL)$, $\Im(\apaR)$,
  $\Re(\apaL)$, and $\Im(\apeL)$ to zero.}

\subsection{Fit Quality}

The effect of adding the fourth symmetry constraint was tested, by
comparing ensembles of three- and four-symmetry fits. The two
ensembles were generated with the same random seed values so that the
ensemble of input data sets was the same for the two approaches. The
fixed spin-amplitude components were chosen to be $\Im(\apaL)$,
$\Im(\apaR)$, $\Re(\apaL)$, and in the case of the four-symmetry fit
also $\Im(\apeL)$.  The amplitudes were still normalised relative to
$\Re({\apaL})$ at $X_{0} = 3.5\gevgev$, however the fits were
performed in the range $\qsq \in [2.5,6]\gevgev$ to avoid the
non-polynomial features seen in the spin-amplitude components, such
as shown in \fig{fig:imA0L-T}.

The sensitivities found for the angular observables are poorer than
those presented in~\cite{Egede:2008uy}, due to the decreased signal
statistics in the reduced \qsq window, however, it is interesting to
compare the performance of the two fitting methods. A histogram of the
NLL of each fit is shown in \fig{fig:10fb_nll-34}. The ensemble of
three-symmetry fits (hatched) and four-symmetry fits (solid) can be
seen. The ensemble of input data sets is slightly different in each
case due to a small number of failed computing jobs, but the output
distributions look very similar. This shows that the depth of the
minima found is approximately the same for the three- and
four-symmetry fits. We can also introduce a global correlation factor
$G_{C}$, which is the unsigned mean of the individual global
correlation coefficients calculated from the full covariance
matrix. It takes values in the range $G_{C} \in [0,1]$, where zero
shows all variables as completely uncorrelated, and one shows total fit
correlation. It can be seen in \fig{fig:10fb_gcorr-34} that the mean
correlation of the fit is reduced once the fourth symmetry is taken
into account. There are less outliers at very low $G_{C}$ and the
distribution appears more Gaussian, indicating an increase in fit
stability has been achieved. The convergence of the fit starting from
arbitrary initial parameters has also much improved.

\DOUBLEFIGURE{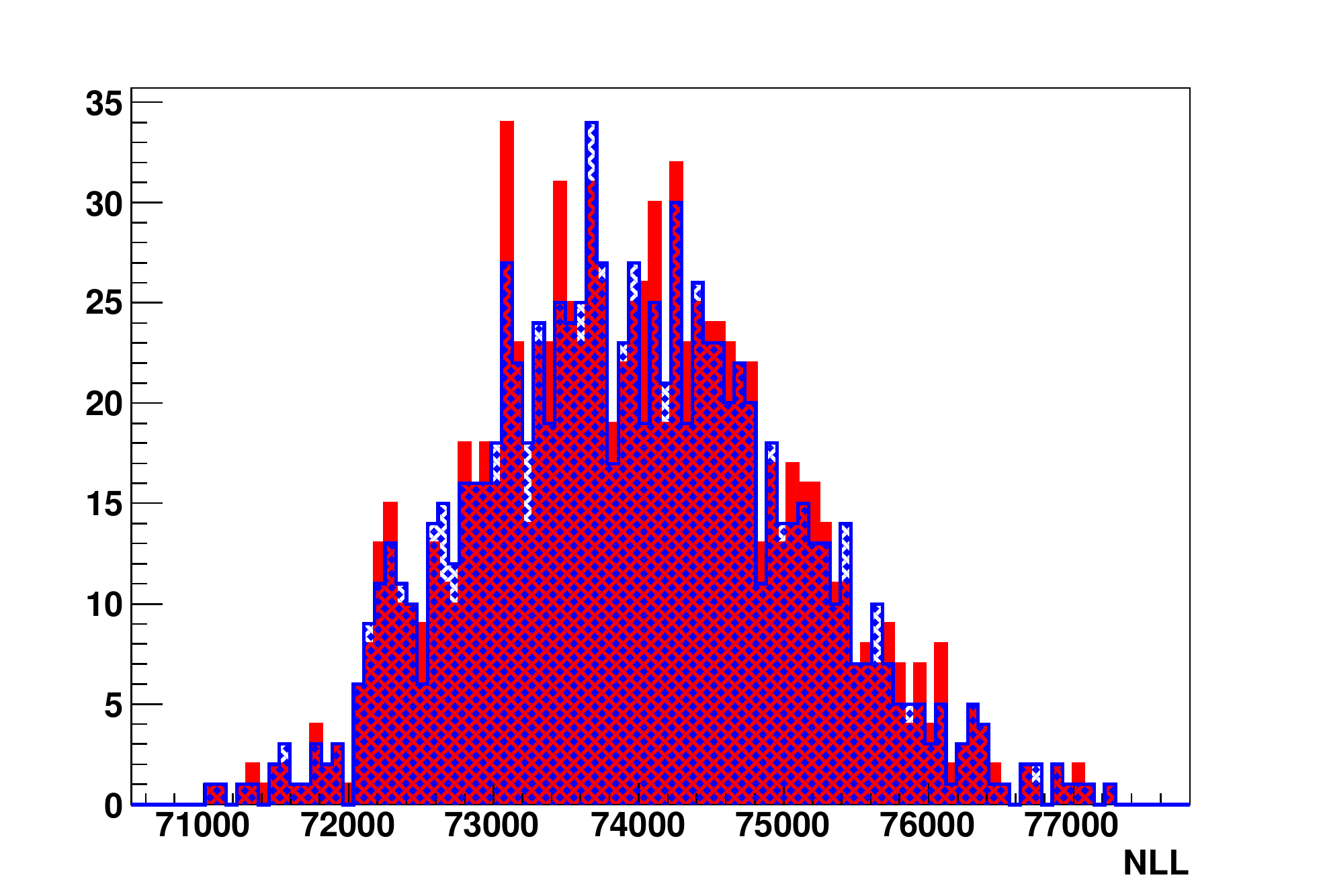,width=0.48\textwidth}{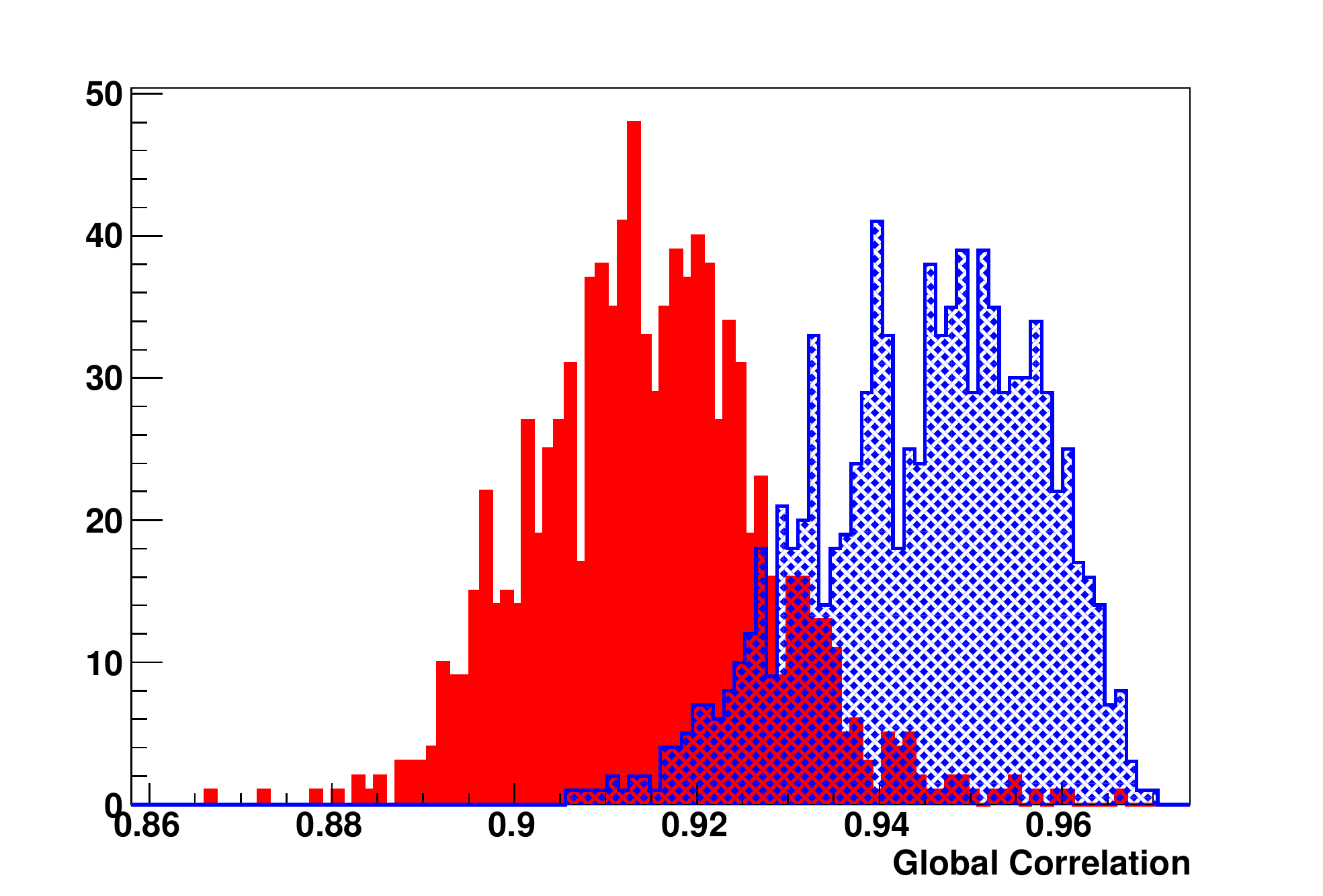,width=0.48\textwidth}{ \label{fig:10fb_nll-34}The
  negative log--likelihood factor for the three-symmetry (blue
  hatched) and four-symmetry (red solid) ensembles of fits to 10\invfb
  toy data sets of \lhcb data, assuming the SM and with $\qsq \in
  [2.5,6]\gevgev$.}{\label{fig:10fb_gcorr-34}The global
  correlation factor for the three-symmetry and four-symmetry
  ensembles of fits to 10\invfb toy data sets of \lhcb data,
  assuming the SM and with $\qsq \in [2.5,6]\gevgev$. The colour
  scheme is the same as in \protect\fig{fig:10fb_nll-34}.}

\fig{fig:10fb_at3-34sym} shows the estimated experimental
sensitivities found for the theoretically clean observable \AT{3} in
the range $\qsq \in [2.5,6]\gevgev$, with and without the fourth
symmetry constraint. The fits are for 10\invfb of \lhcb integrated
luminosity assuming the SM. As might be expected from
\fig{fig:10fb_nll-34}, there is little difference in the estimated
experimental resolutions seen. The same conclusion is reached when
inspecting other observables. 

\FIGURE{
\includegraphics[width=0.48\textwidth]{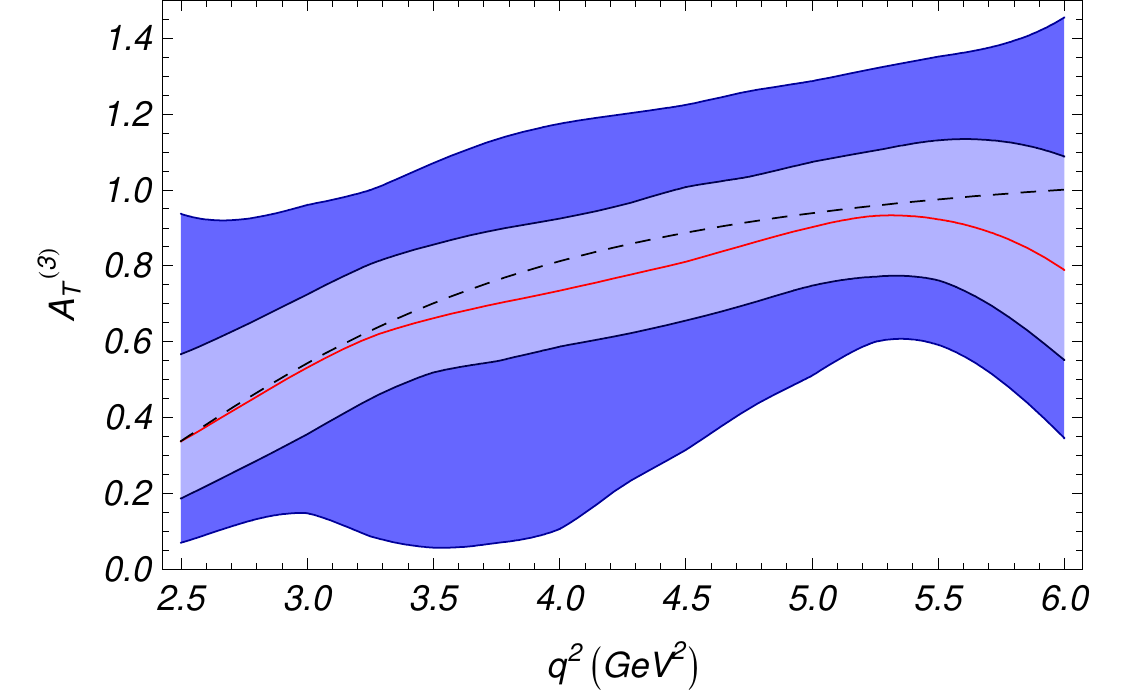}
\includegraphics[width=0.48\textwidth]{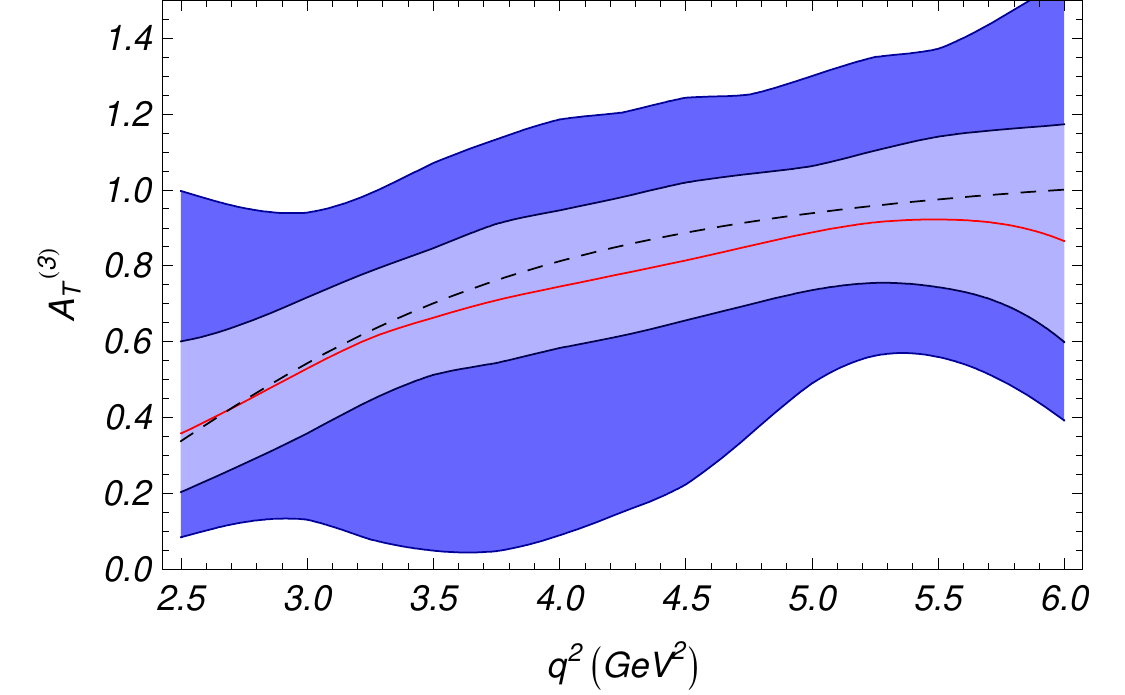}
\caption{ \label{fig:10fb_at3-34sym} One and two $\sigma$ contours of
  estimated experimental sensitivity to the theoretically clean
  observable \AT3 with full-angular fit to 10\invfb of \lhcb data
  assuming the SM. The results of the three-symmetry fit are shown on
  the left, and the four-symmetry fit on the right. The fits were
  performed in the range $\qsq \in [2.5,6]\gevgev$.}
}

\subsection{Discussion}
The discovery of a fourth symmetry in the massless leptons limit of
the full-angular distribution of \BdbKsmm requires that the
experimental analysis proposed in~\cite{Egede:2008uy} be
re-evaluated. The previous analysis used three of the four available
symmetry constraints to perform a fit, that was, in principle,
under-constrained, by parametrising the real and imaginary parts of
the \Kstarz spin amplitudes as second-order polynomials. The
invariance of the observables under all four symmetries, and the
freedom to take arbitrary values of the \(\tilde{\theta}\) rotation angle,
allowed the fits to converge and produce correct output, but
introduced a subtle parametrisation bias. As the observables are
\emph{by definition} invariant to all the symmetries, the estimated
experimental sensitivities are the same for the two methods. This has been
demonstrated in this section. However, the need for the development of a new fitting method,
so that the full experimental statistics available in $\qsq \in [1,6]\gevgev$ can be used, is now clear.
The sensitivities found will be similar to those estimated in \rf{Egede:2008uy} and in this paper for the \CP asymmetries, but with improved fit stability.

\section{Analysis of  \CP-violating observables}
\label{sec:CPviolating}

In~\cite{Kruger:1999xa,Hewett:2004tv}, it was shown that eight
\CP-violating observables can be constructed by combining the
differential decay rates of $d \Gamma(\BdbKsll)$ and $d
\overline{\Gamma}(\BdKsll)$. In this section we analyse the
theoretical and experimental uncertainties of those observables in
order to judge the NP sensitivity of such \CP-violating observables.

\subsection{Preliminaries}
The corresponding decay rate for the \CP-conjugated decay mode \BdKsll
is given by 
\begin{equation}
  \frac{d^4 \overline{\Gamma}}{dq^2\, d\cos\theta_l\, d\cos\theta_{K}\, d\phi} =
   \frac{9}{32\pi} \bar{J}(q^2, \theta_l, \theta_{K}, \phi)\,.
\end{equation} 
As shown in \cite{Kruger:1999xa}, the corresponding functions 
$\bar{J}_i( q^2, \theta_l, \theta_K, \phi)$  are connected to functions $J_i$ in the following way:
\begin{equation} 
J_{1,2,3,4,7} \rightarrow \bar{J}_{1,2,3,4,7}, \,\,\,\,\, J_{5,6,8,9} \rightarrow - \bar{J}_{5,6,8,9} \,,
\label{Jtrafos}
\end{equation}
where $\bar{J}_i$ equals $J_i$ with all weak phases conjugated.

Besides  the \CP asymmetry in the dilepton mass distribution,  there are several \CP-violating observables
in the angular distribution.  The latter are sensitive to \CP-violating effects as differences between 
the angular coefficient functions, $J_i - \bar{J}_i$.  As was discussed   in~\cite{Kruger:1999xa,Hewett:2004tv}, and more recently  in~\cite{Bobeth:2008ij}, those \CP asymmetries are all very small in the SM;
they  originate from the  small \CP-violating imaginary part of $\lambda_u = (V_{ub}V^*_{us})/(V_{tb} V^*_{ts})$. This weak phase present in the Wilson coefficient $\Ceff9$ is doubly-Cabibbo suppressed  and  further suppressed by the ratio of the Wilson coefficients $(3 \C1 +\C2) / \C9 \approx 0.085$. 

Moreover, it is important to note \cite{Hewett:2004tv,Bobeth:2008ij}
that the \CP asymmetries corresponding to $J_{7,8,9}$ are odd under
the transformation $\phi \rightarrow - \phi$ and thus, these
asymmetries are T-odd (T transformation reverses all particle momenta
and particle spins) while the other angular \CP asymmetries are
T-even.  T-odd \CP asymmetries are favoured because they involve the
combination $\cos(\delta \theta)\sin(\delta \phi_W)$ of the strong
and weak phase differences \cite{Hewett:2004tv,Bobeth:2008ij}, thus,
they are still large in spite of small strong phases as predicted for
example within the QCDF/SCET approach.  In contrast, T-even \CP
asymmetries involve the combination $\sin(\delta \theta)\cos
(\delta \phi_W)$~\cite{Hewett:2004tv,Bobeth:2008ij}\footnote{We note
  here that this specific behaviour of T-odd and T-even observables
  was shown in many examples of T-odd \CP asymmetries
  (see~\cite{Valencia:1994zi} and references therein) but a general
  proof of this statement is still missing to our knowledge.}.

Another remark is that the \CP asymmetries related to $J_{5,6,8,9}$
can be extracted from $( d \Gamma + d \overline{\Gamma} )$ due to the
property Eq.~(\ref{Jtrafos}), and thus can be determined for an untagged
equal mixture of \B and \Bdb mesons.  This is important for the
decay modes $B^0_d \rightarrow K^{*0} ( \rightarrow K^0 \pi^0 )
\ell^+\ell^-$ and $B_s \rightarrow \phi (\rightarrow K^+K^-)
\ell^+\ell^-$ but it is less relevant for the self-tagging mode $B_d
\rightarrow K^{*0} (\rightarrow K^+\pi^-)\ell^+\ell^-$.

Recently, a QCDf/SCET analysis of the angular \CP-violating
observables, based on the NLO results in~\cite{Beneke:2001at,
  Beneke:2004dp}, was presented for the first
time~\cite{Bobeth:2008ij}.  The NLO corrections are shown to be
sizable. The crucial impact of the NLO analysis is that the scale
dependence gets reduced to the $10\%$ level for most of the \CP
asymmetries. However, for some of them, which essentially start with a
nontrivial NLO contribution, there is a significantly larger scale
dependence.  The $q^2$-integrated SM predictions are all shown to be
below the $10^{-2}$ level due to the small weak phase as mentioned
above. The uncertainties due to the form factors, the scale
dependence, and the uncertainty due to CKM parameters are identified
as the main sources of SM errors~\cite{Bobeth:2008ij}.

\subsection{Phenomenological analysis}
The NP sensitivity of \CP-violating observables in the mode \BdbKsll
was discussed in a model-independent way~\cite{Bobeth:2008ij} and also
in various popular concrete NP models~\cite{Altmannshofer:2008dz}.  It
was found that the NP contributions to the phases of the Wilson
coefficients $\C7$, $\C9$, and $\C{10}$ and of their chiral
counterparts drastically enhance such \CP-violating observables, while
presently most of those phases are very weakly constrained.  It was
claimed that these observables offer clean signals of NP
contributions.

However, the NP reach of such observables can only be judged with a
{\it complete} analysis of the theoretical and experimental
uncertainties.  To the very detailed analyses
in~\cite{Bobeth:2008ij,Altmannshofer:2008dz} we add the following
points:
\begin{itemize}
\item We redefine the various \CP asymmetries following the general
  method presented in our previous paper \cite{Egede:2008uy}: an
  appropriate normalisation of the \CP asymmetries almost eliminates
  any uncertainties due to the soft form factors which is one of the
  major sources of errors in the SM prediction.
\item We explore the effect of the possible $\Lambda_{\rm QCD}/m_b$ corrections
  and make the uncertainty due to those unknown $\Lambda_{\rm QCD}/m_b$
  corrections manifest in our analysis within the SM and NP scenarios.
\item We investigate the experimental sensitivity of the angular \CP
  asymmetries using a toy Monte Carlo model and estimate the
  statistical uncertainty of the observables with statistics
  corresponding to five years of nominal running at LHCb ($10\invfb$)
  using a full angular fit method.
\end{itemize}
We discuss these issues by example of  the two  angular asymmetries 
corresponding to the angular coefficient functions $J_{6 s}$ and $J_8$;
\begin{equation}
 A_{6s}= \frac{J_{6s} - \bar J_{6s}}{d(\Gamma+\overline\Gamma)/dq^2}, \quad
  A_{8} =\frac{J_{8} - \bar J_{8} } {d(\Gamma+\overline\Gamma)/dq^2} \, .
\label{definitions}
\end{equation}
Within the SM the first \CP asymmetry related to $J_{6s}$ turns out
to be the well-known forward-backward \CP asymmetry which was proposed
in~\cite{Buchalla:2000sk,Kruger:2000zg}.

As a first step we redefine the two \CP observables. We  make  sure that the form factor dependence cancels out at the LO level by using an appropriate normalisation: 
\begin{equation}
 A^{V2s}_{6s}= \frac{ J_{6s} - \bar J_{6s} }{J_{2s} +  \bar J_{2s}}, \quad
  A^V_{8} = \frac{ J_{8} - \bar J_{8}}{J_{8} + \bar J_{8}} \, .
\label{newdefinitions}
\end{equation}
\FIGURE[t]{
\includegraphics[width=.48\textwidth]{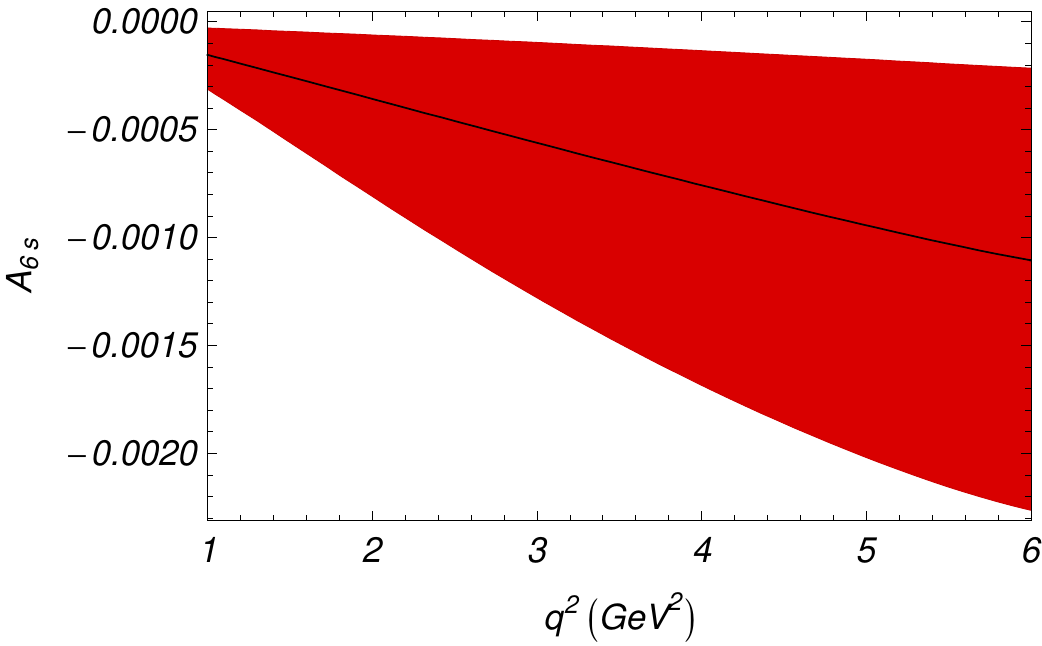}
\includegraphics[width=.48\textwidth]{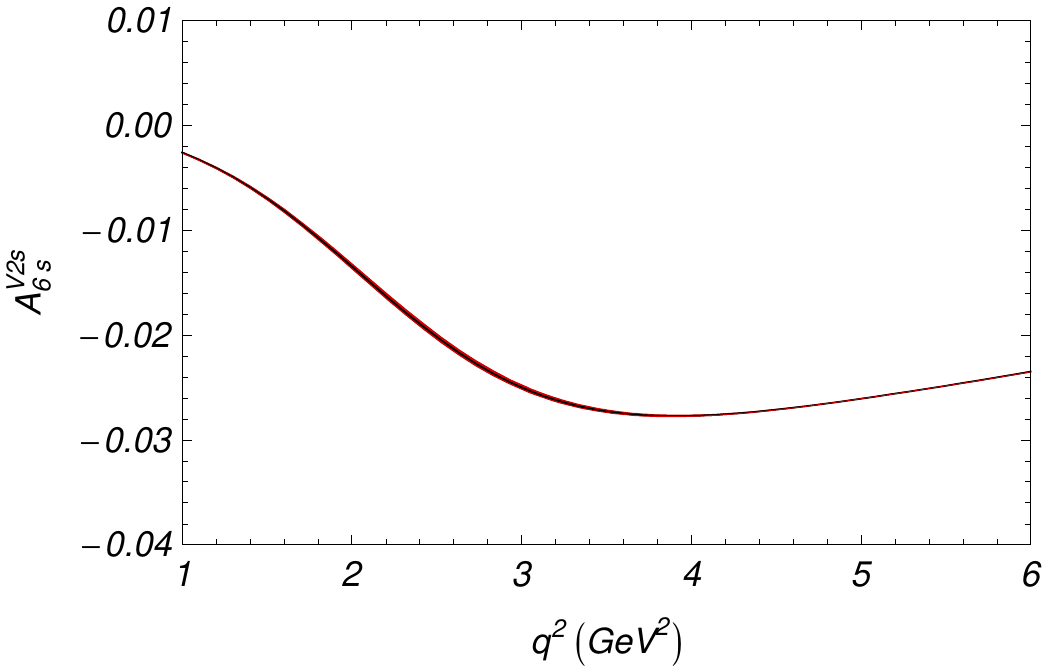}
\caption{SM prediction of the \CP-violating observables $A_{6s}$ (left) and
  $A_{6s}^{V2s}$ (right) as function of the squared lepton mass with
  uncertainty due to the soft form factors only.  Notice  the difference in scale and the difference in relative error in the two figures.
  \label{formfactorA62s}}
}
\FIGURE[t]{
\includegraphics[width=.48\textwidth]{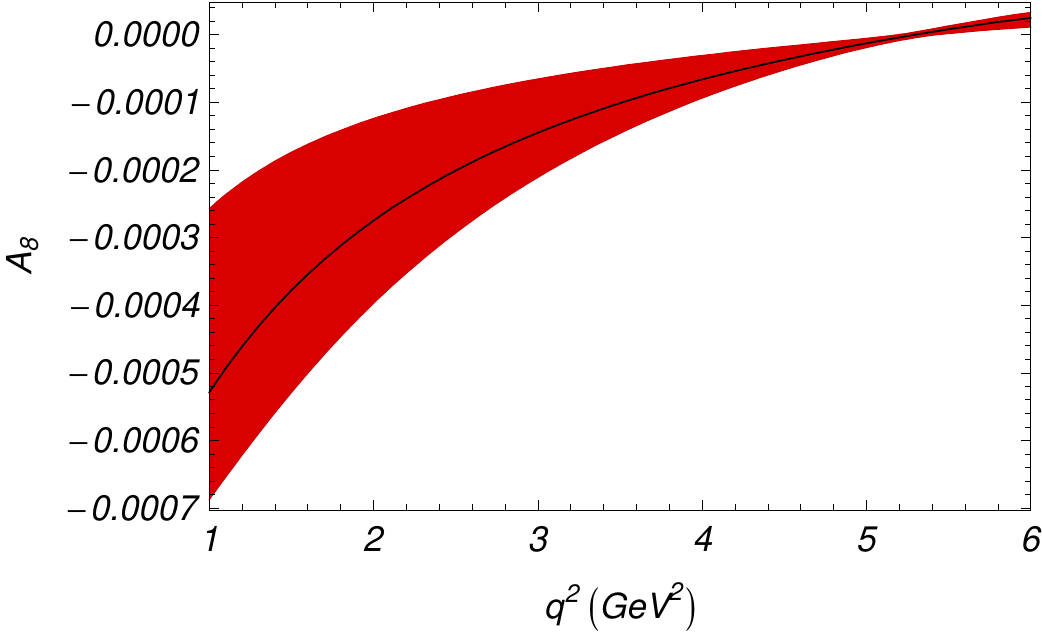}
\includegraphics[width=.48\textwidth]{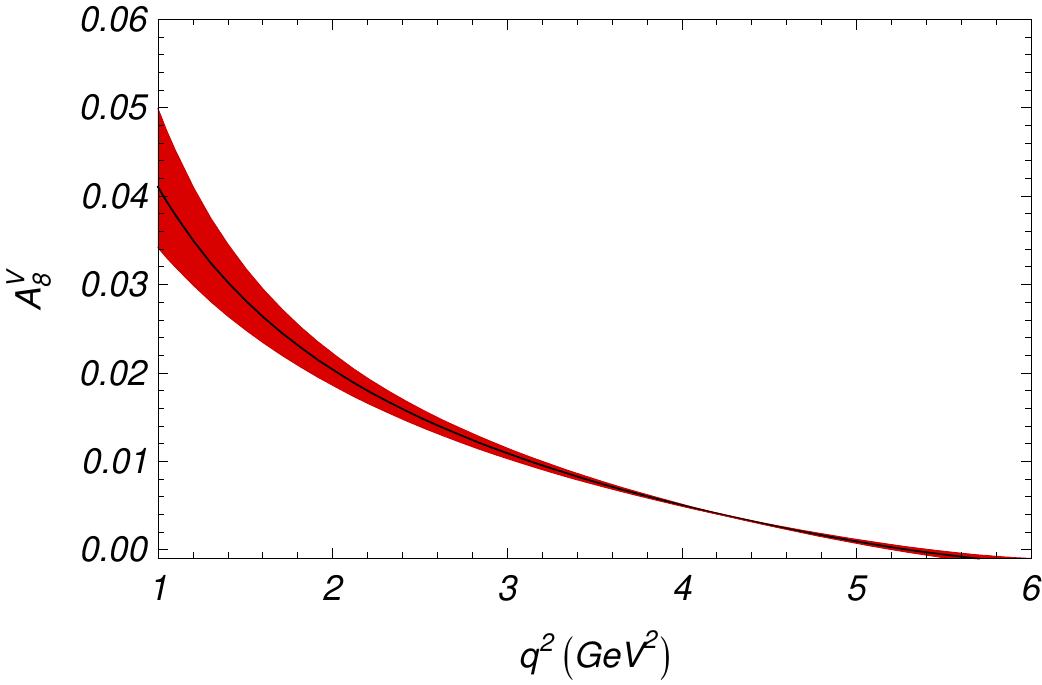}
\caption{SM
  prediction of the \CP-violating observables $A_{8}$
  (left) and $A_{8}^V$ (right) with uncertainty due to
  the soft form factors only. Notice the difference in scale of the two figures.
  \label{formfactorA8}}
}
The $J_i$ are bilinear in the $K^*$ spin amplitudes, so it is clear
from the LO formulae Eq.~(\ref{SCETKspin}) that, following the
strategy of~\cite{Egede:2008uy}, any form factor dependence at this
order cancels out in both observables. We note that $J_{2s}$ has the
same form factor dependence as $J_{6s}$ but has larger absolute values
over the dilepton mass spectrum that stabilises the quantity.  In
Fig.~\ref{formfactorA62s} the uncertainty due to the form factor
dependence is estimated in a conservative way  (see Appendix~\ref{sec:InputAppendix})  for $A_{6s}$ defined in
Eq.~(\ref{definitions}) and for $A_{6s}^V$ defined in
Eq.~(\ref{newdefinitions}).  Comparing the plots, one sees that with
the appropriate normalisation, this main source of hadronic
uncertainties gets almost eliminated.  The leftover uncertainty enters
through the form factor dependence of the NLO contribution.
Fig.~\ref{formfactorA8} shows the analogous results for the observable
$A_8^V$.  

\enlargethispage{\baselineskip}
In the second step we make the possible $\Lambda_{\rm QCD}/m_b$ corrections
manifest in our final results by using the procedure described in
Sec.~\ref{sec:LambdaOverMb}. It turns out that in spite of this very
conservative ansatz for the possible power corrections (we neglect for
example any kind of correlations between such corrections in the
various spin amplitudes), the impact of those corrections is smaller
than the SM uncertainty in case of the two observables $A_{6s}^V$ and
$A_8^V$. In the left plot of Fig.~\ref{SMerrorA6s} the SM error is
given, including uncertainties due to the scale dependence and input
parameters and the spurious error due to the form factors.  In the
right plot the estimated power corrections are given, which in case of
the \CP-violating observable $A_{6s}^V$ are significantly smaller than
the combined uncertainty due to scale and input parameters.
Fig.~\ref{SMerrorA8} shows the same feature for the \CP-violating
observable $A_8^V$.  This result is in contrast to the one for
\CP-averaged angular observables discussed in~\cite{Egede:2008uy},
where the estimated power corrections always represent the dominant
error. The reason for this specific feature is the
smallness of the weak phase in the SM. Thus, one expects that the
impact of power corrections will be significantly larger when NP
scenarios with new \CP phases are considered (see below).
\FIGURE{
\includegraphics[width=.48\textwidth]{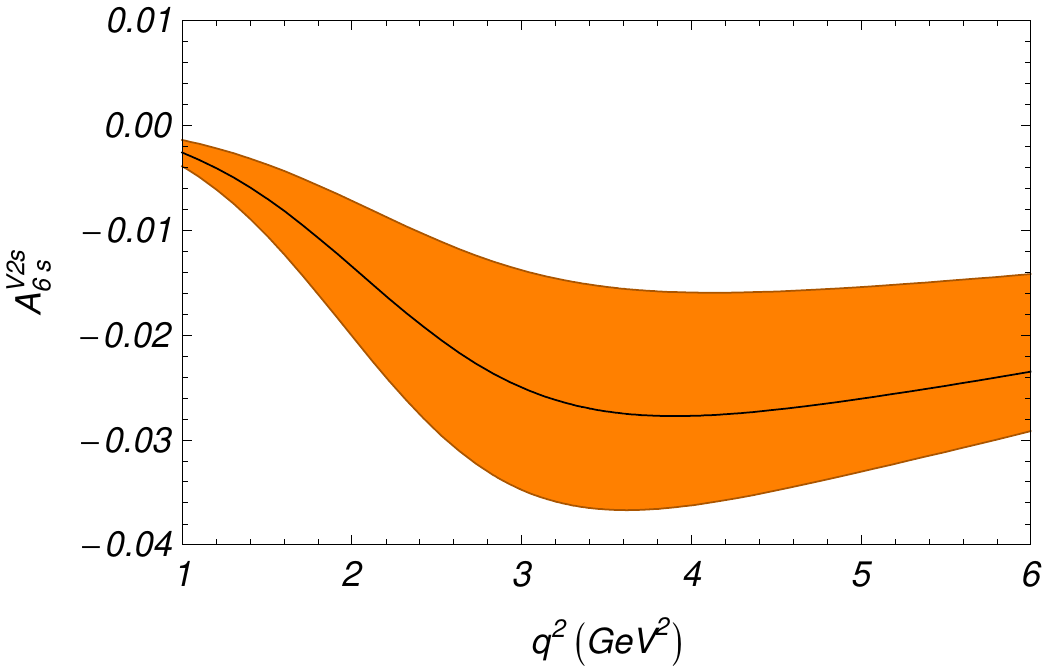}
\includegraphics[width=.48\textwidth]{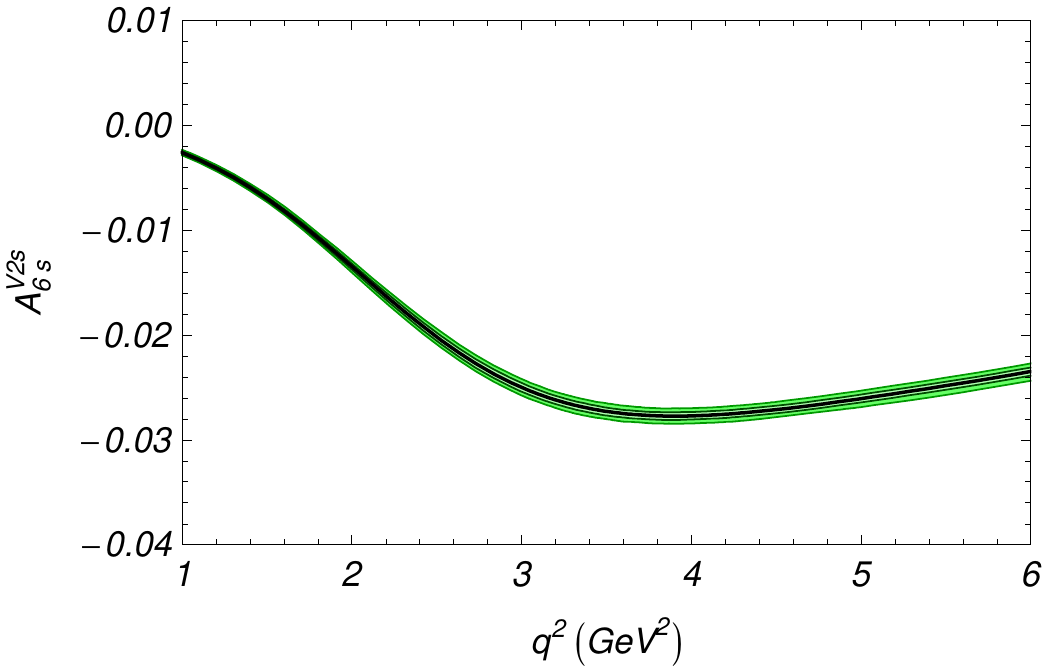}
\caption{\label{SMerrorA6s}SM
  uncertainty in $A_{6s}^{V2s}$ (left) and estimate of uncertainty due
  to $\Lambda_{\rm QCD}/m_b$ corrections with $C_{1,2} = 10\%$ (right).}
}
\FIGURE{
\includegraphics[width=.48\textwidth]{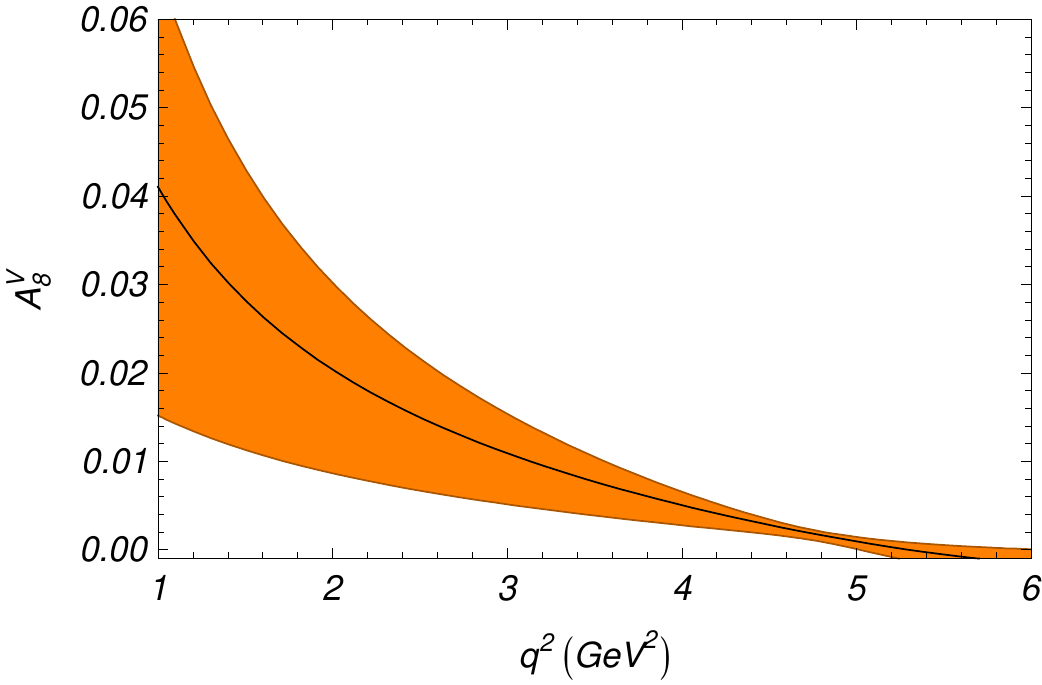}
\includegraphics[width=.48\textwidth]{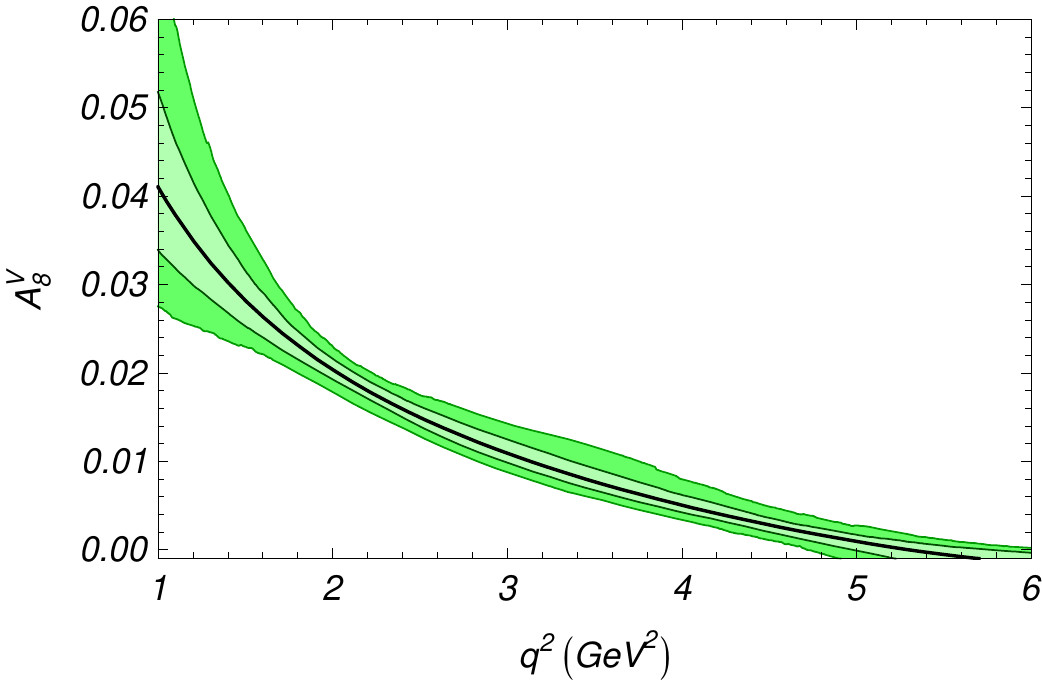}
\caption{\label{SMerrorA8}SM
  uncertainty in $A_{8}^{V}$ (left) and estimate of uncertainty due to
  $\Lambda_{\rm QCD}/m_b$ corrections (right, light grey (green) corresponds to
  $C_{1,2} = 5\%$, dark grey (green) to $C_{1,2} = 10\%$).}
}

In the third step we consider various NP scenarios. Here we follow the
model-independent constraints derived in~\cite{Bobeth:2008ij} assuming
only one NP Wilson coefficient being nonzero. We consider three
different NP benchmarks scenarios of this kind:
\begin{enumerate} 
\item  $| \C9^{\scriptscriptstyle{\rm NP}} | = 2$ and $\phi_9^{\scriptscriptstyle{\rm NP}} = \frac{\pi}{8}, \frac{\pi}{2}, \pi$ (Red);
\item  $| \C{10}^{\scriptscriptstyle{\rm NP}} | = 1.5$ and $\phi_{10}^{\scriptscriptstyle{\rm NP}} =  \frac{\pi}{8}, \frac{\pi}{2}, \pi$ (Grey);
\item   $| \Cp{10}| = 3$ and $\phi_{10}^{'} = \frac{\pi}{8}, \frac{\pi}{2}, \pi$ (Blue); 
\end{enumerate} 
where the colours refer to the ones used in the following figures. The
absolute values of the Wilson coefficients are chosen in such a way
that the model-independent analysis, assuming {\it one} nontrivial NP
Wilson coefficient acting at a time, does not give any bound on the
corresponding NP phase.

\FIGURE{
\includegraphics[width=.48\textwidth]{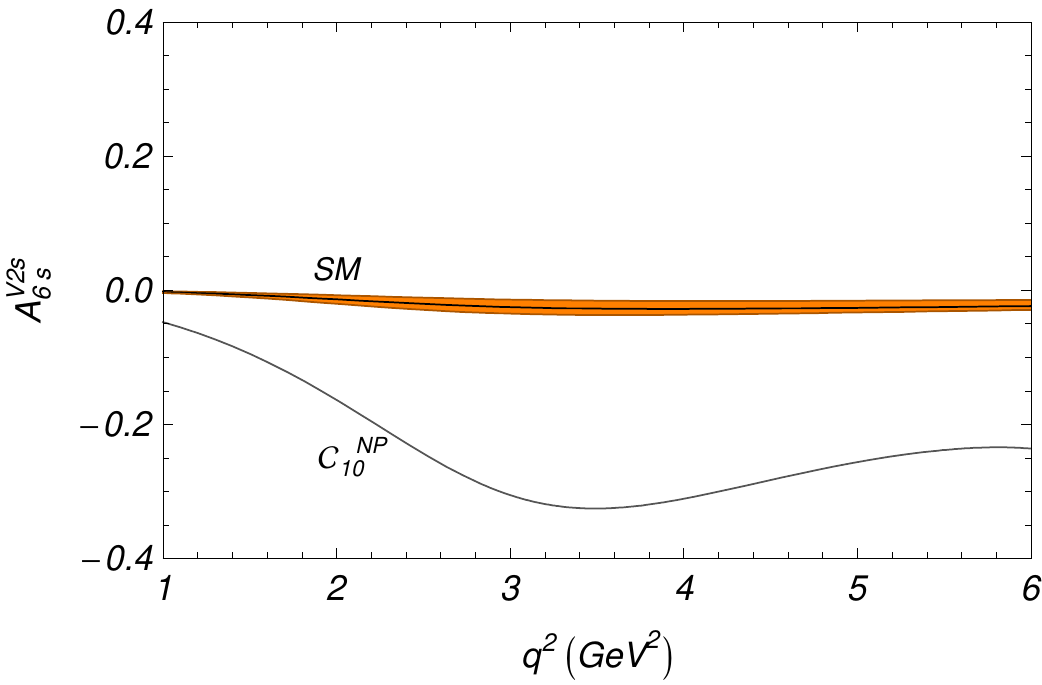}
\includegraphics[width=.48\textwidth]{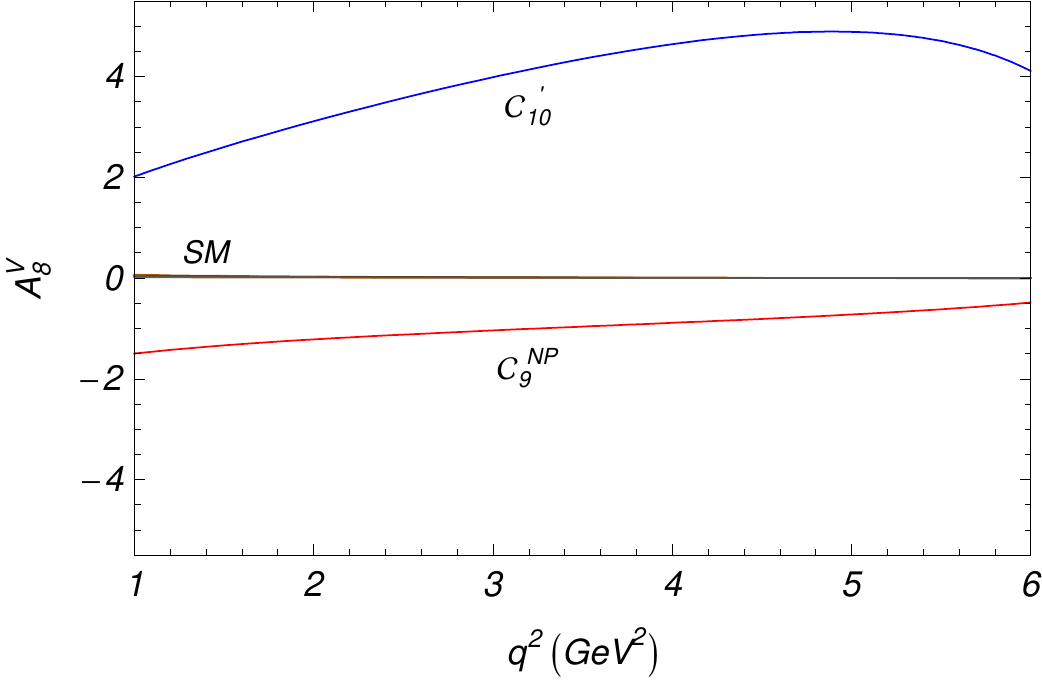}
\caption{\label{fig:CPVnewphysics}NP
  scenarios, assuming one nontrivial NP Wilson coefficient at a time,
  next to SM prediction for $A_{6s}^{V2s}$ (left) and $A_8^V$ (right),
  for concrete values see text.}
}
Fig.~\ref{fig:CPVnewphysics} shows our two observables in the three
scenarios with the phase value $\frac{\pi}{8}$: the \CP-violating observable
$A_{6s}^V$ might separate a NP scenario (2), while the central values
of scenarios (1) and (3) are very close to the SM. Moreover observable
$A_8^V$ seems to be suited to separate scenarios (1) and (3) from the
SM.

However, to judge the NP reach we need a complete error analysis
within the three NP scenarios. As shown in Sec.~\ref{sec:LambdaOverMb}
we now work with three weak sub-amplitudes in which possible power
corrections are varied independently. The plots in
Figs.~\ref{lambdaA6s} and \ref{lambdaA8} show that the possible
$\Lambda_{\rm QCD}/m_b$ corrections have a much larger impact on our two
observables in the NP scenarios than in the SM and become the
dominating theoretical uncertainty.  We also get significantly larger
possible $\Lambda_{\rm QCD}/m_b$ corrections when changing the value of the new
weak phase from $\frac{\pi}{8}$ to $\frac{\pi}{2}$.  Regarding even
larger phase values, we note here that the NP effects drastically
decrease again when phase values around $\pi$ are chosen as
expected. Nevertheless, in view of the theoretical $\Lambda_{\rm QCD}/m_b$
uncertainties only, the two \CP-violating observables could
discriminate some specific NP scenarios with new \CP phase of order
$\frac{\pi}{8}$ or $\frac{\pi}{2}$ from the SM; in case of
$A_{6s}^{V2s}$ NP scenario 2, in case of $A_{8}^{V}$ NP scenario 3 and
possibly 1.  
\FIGURE{
\includegraphics[width=.48\textwidth]{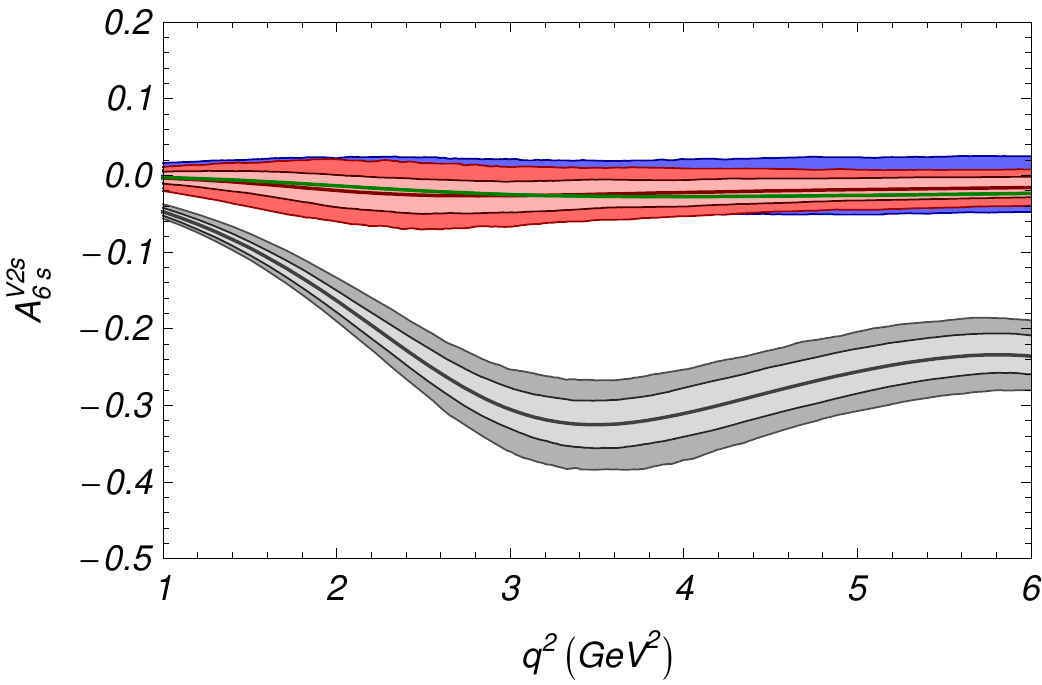}
\includegraphics[width=.48\textwidth]{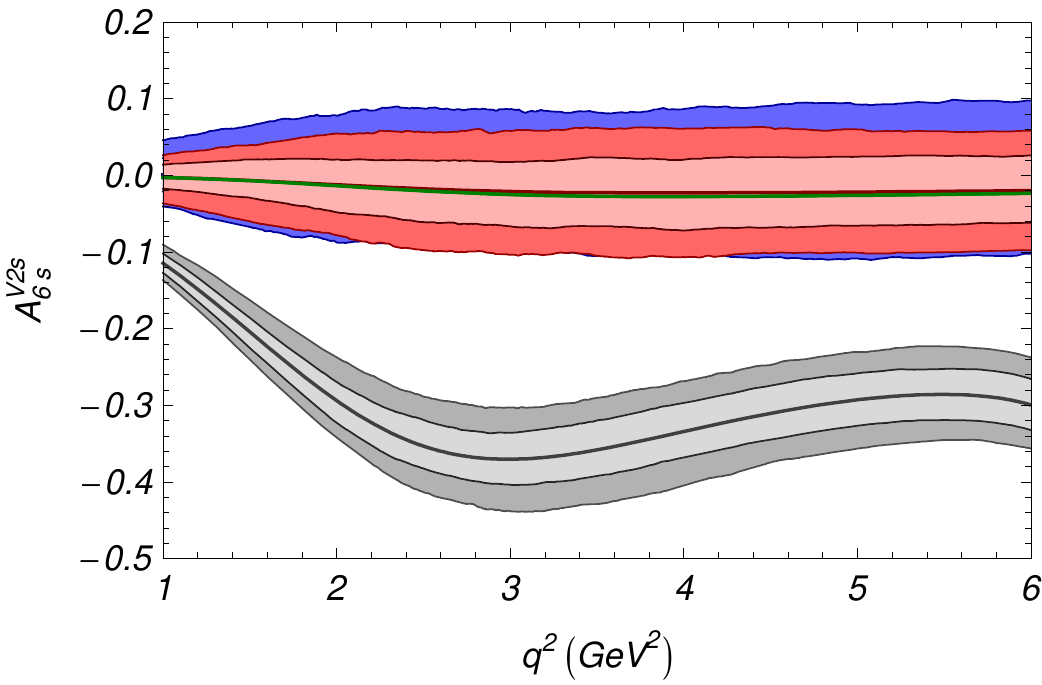}
\caption{\label{lambdaA6s}$A_{6s}^{V2s}$:
  Estimate of uncertainty due to $\Lambda_{\rm QCD}/m_b$ corrections within NP
  scenarios as in previous figure with phases $\phi_i = \frac{\pi}{8}$
  (left) and $\phi_i=\frac{\pi}{2}$ (right).}
}
\FIGURE{
\includegraphics[width=.48\textwidth]{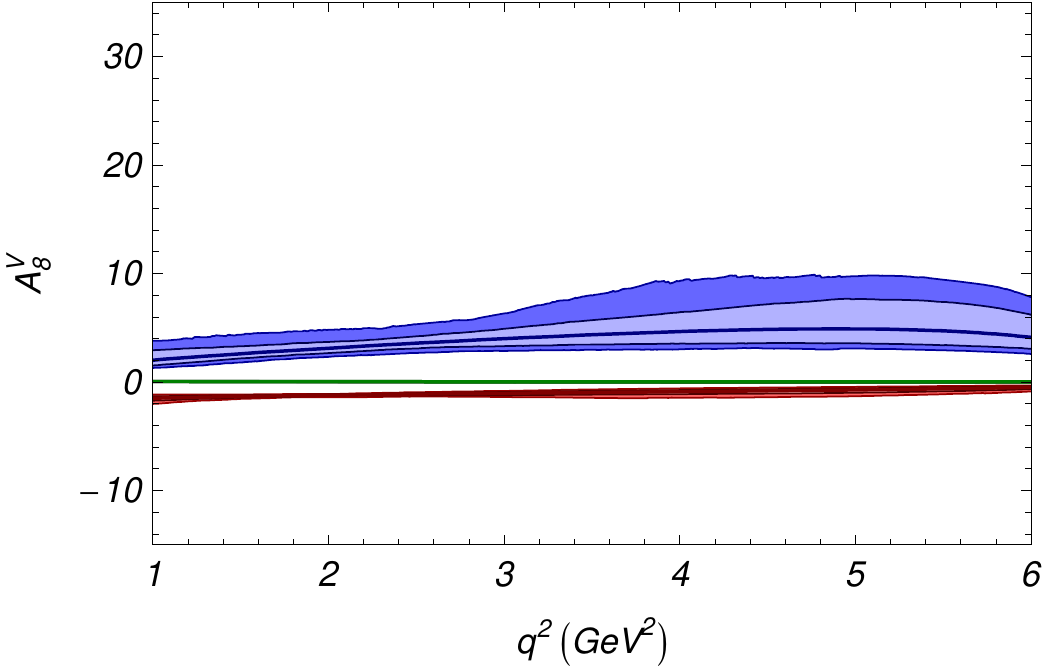}
\includegraphics[width=.48\textwidth]{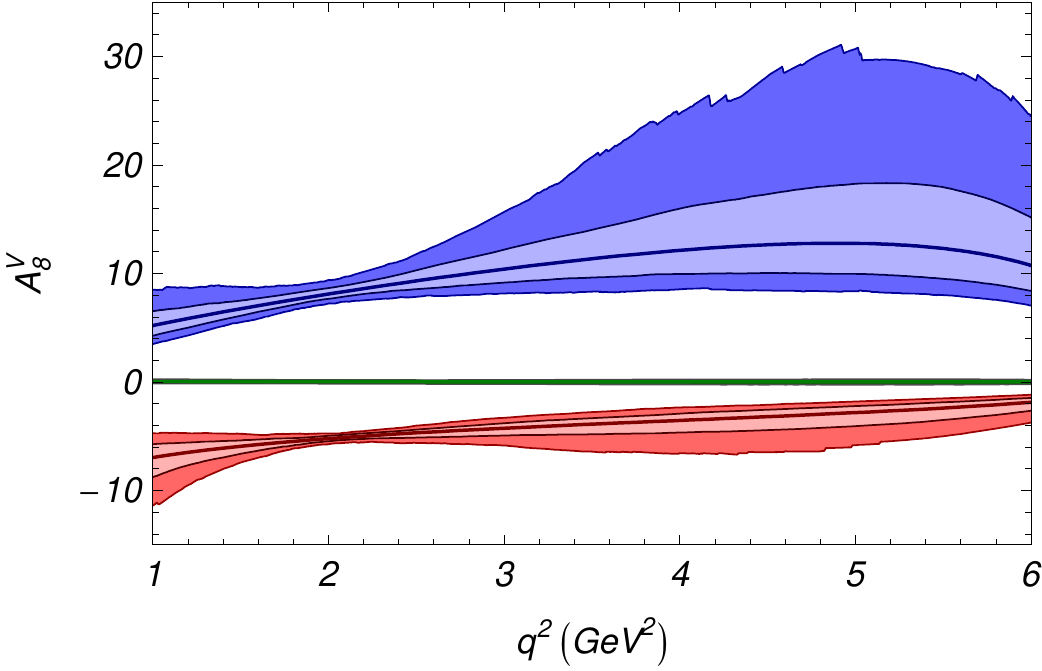}
\caption{\label{lambdaA8}$A_{8}^{V}$:
  Estimate of uncertainty due to $\Lambda_{\rm QCD}/m_b$ corrections within NP
  scenarios as in previous figure with phases $\phi_i = \frac{\pi}{8}$
  (left) and $\phi_i=\frac{\pi}{2}$ (right).} 
}

One should also consider the additional theoretical
uncertainties due to scale dependence, input parameters and soft form
factor dependencies within the NP scenarios. Those additional
theoretical uncertainties are sizable and of the same order as the
ones due to $\Lambda_{\rm QCD}/m_b$ corrections: they are shown in the left
plots in Figs.~\ref{theorytotalexA6s} and \ref{theorytotalexA8} as
orange bands overlaying the {\bf total} errors bars including also the
$\Lambda_{\rm QCD}/m_b$ corrections.
\FIGURE{
\includegraphics[width=.48\textwidth]{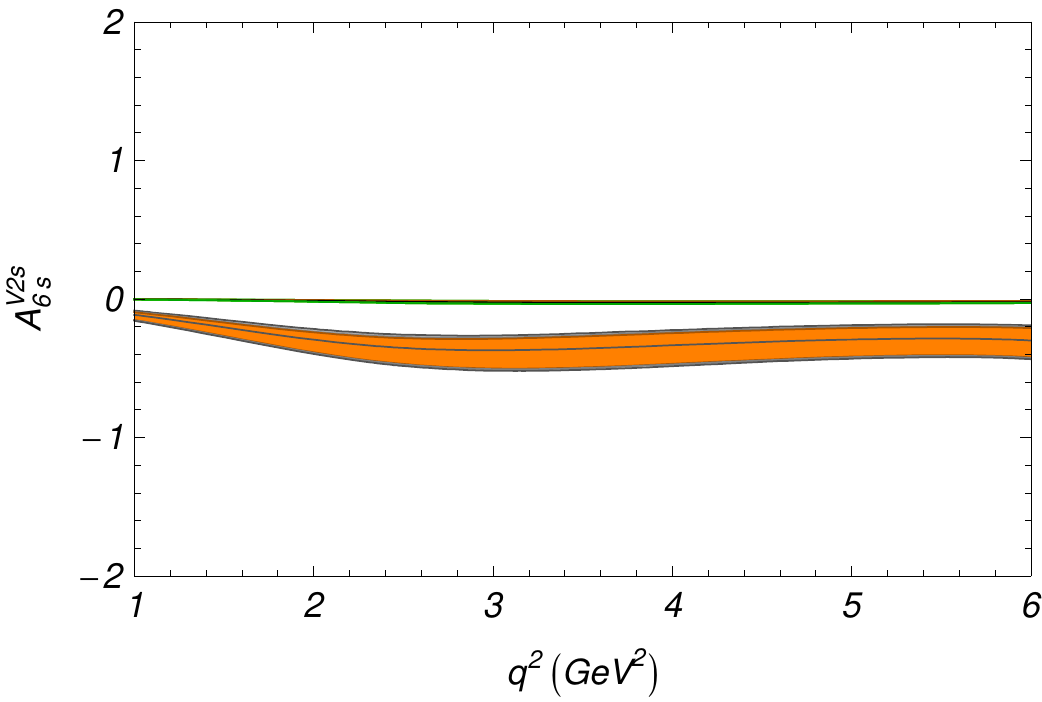}
\includegraphics[width=.48\textwidth]{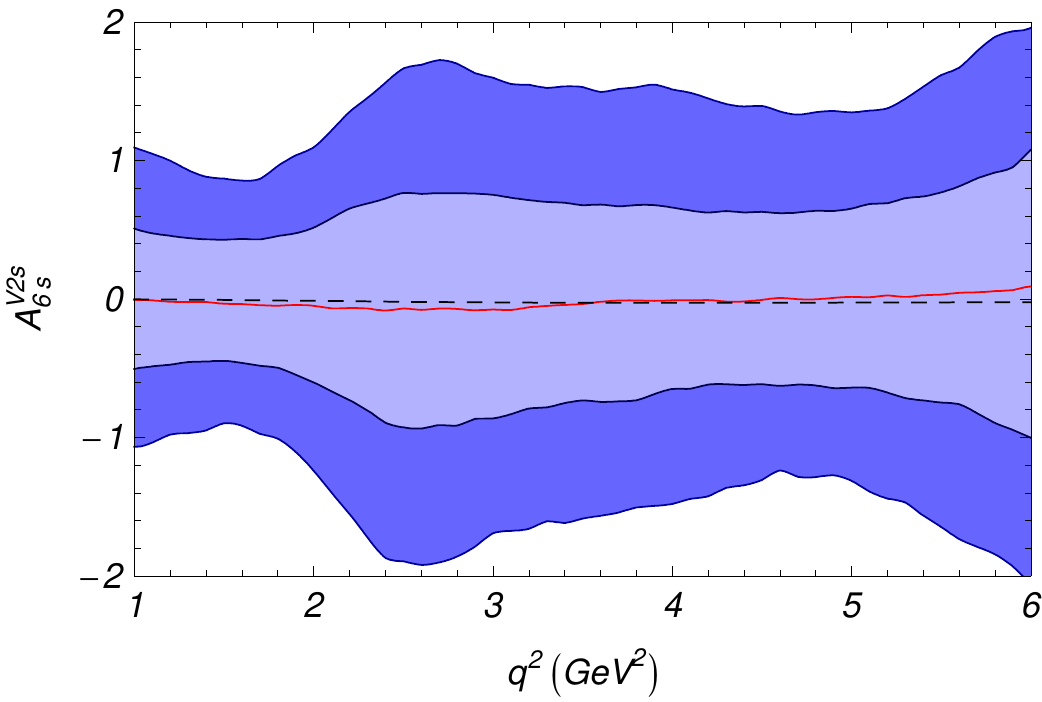}
\caption{\label{theorytotalexA6s}$A_{6s}^{V2s}$:
  Estimate of uncertainty due to $\Lambda_{\rm QCD}/m_b$ corrections (grey
  bands) in NP scenario 2,  $|\C{10}^{\scriptscriptstyle{\rm NP}}|=1.5 $ and $\phi_{10}^{\scriptscriptstyle{\rm NP}} 
  =\frac{\pi}{2}$  with the other theoretical uncertainties overlaid (orange
  bands) and in SM (left) and experimental uncertainty (right).}
}

As the last step, we analyse the experimental sensitivity of the
angular \CP asymmetries using a toy Monte Carlo model.  The right
plots in Figs.~\ref{theorytotalexA6s} and \ref{theorytotalexA8} show
the estimates of the statistical uncertainty of $A_{6s}^V$ and $A_8^V$
with statistics corresponding to five years of nominal running at LHCb
($10\invfb$).  The inner and outer bands correspond to $1\sigma$ and
$2\sigma$ statistical errors.  The plots show that all the NP
benchmarks are within the $1\sigma$ range of the expected experimental
error in case of the observable $A_{6s}^V$, and within the $2\sigma$
range of the experimental error in case of the observable $A_8^V$. We
emphasise that from the experimental point of view the normalisation
is not important when calculating the overall significance because the
overall error is dominated by the error on the numerator. So the
experimental error of the observables $A_{6s}$ and $A_8$ defined in
Eq.~(\ref{definitions}) using the traditional normalisation will be
similarly large to the one of our new observables $A_{6s}^V$ and
$A_8^V$ defined in Eq.~(\ref{newdefinitions}).
\FIGURE{
\includegraphics[width=.48\textwidth]{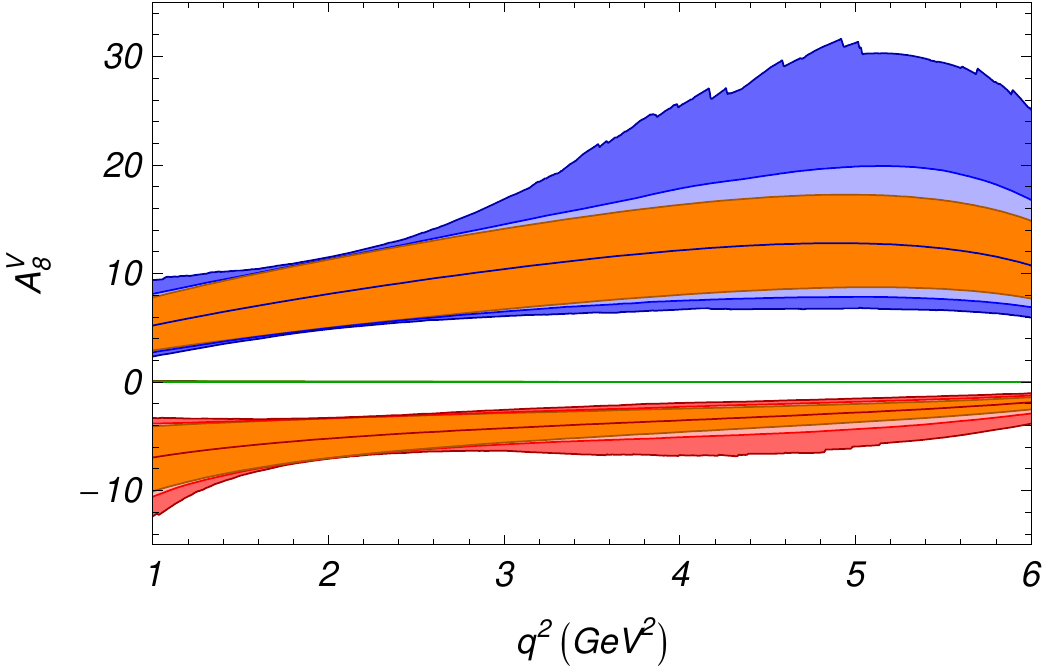}
\includegraphics[width=.48\textwidth]{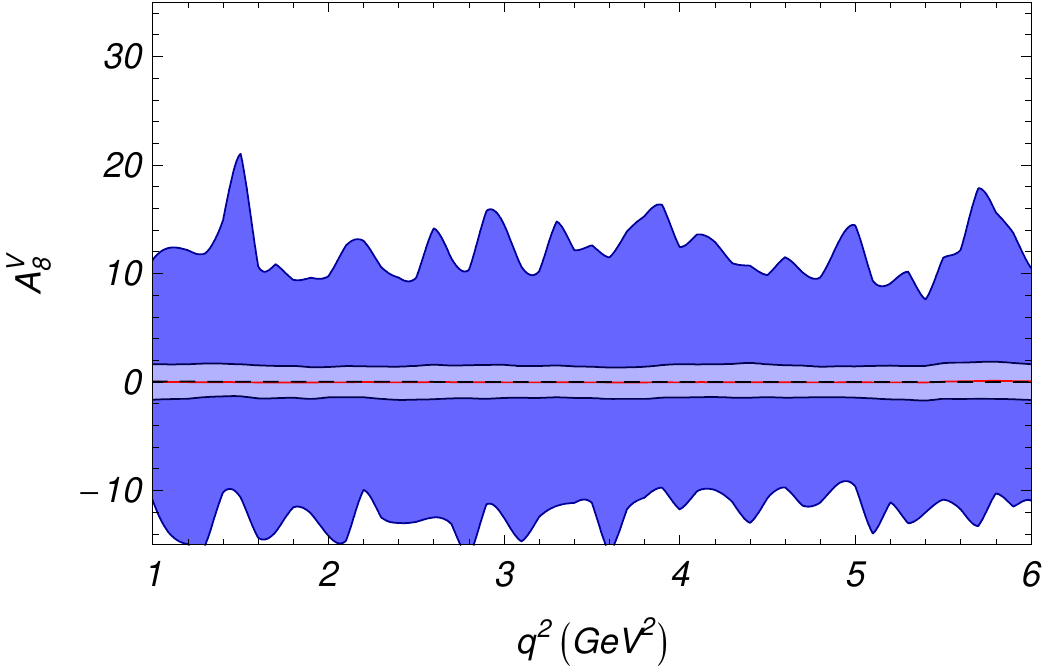}
\caption{\label{theorytotalexA8}$A_8^V$:
  Estimate of uncertainty due to $\Lambda_{\rm QCD}/m_b$ corrections in NP
  scenarios 1 ( $|\C9^{\scriptscriptstyle{\rm NP}}|=2 $, $\phi_{9}^{\scriptscriptstyle{\rm NP}} =\frac{\pi}{2}$, red  bands)
  and 3 ($|\C{10}^{'}|=3 $, $\phi_{10}^{'} =\frac{\pi}{2}$, blue bands)
  with the other theoretical uncertainties overlaid (orange bands) and
  in SM (left) and experimental uncertainty (right).}
}

Our final conclusion is that the possibility to disentangle different
NP scenarios for the \CP-violating observables remains rather
difficult. For the rare decay \BdbKsll, \lhcb has no real sensitivity
for NP phases up to values of $\frac{\pi}{2}$ (and neither up to
values of $\pi$) in the Wilson coefficients $\C9$, $\C{10}$ and their
chiral counterparts. Even Super-\lhcb with $100\invfb$ integrated
luminosity does not improve the situation significantly. This is in
contrast to the \CP-conserving observables presented
in~\cite{Egede:2008uy} and further discussed in the next chapter
which, both from the theoretical and experimental point of view, are
very promising.

\section{Analysis of  \CP-conserving observables}
\label{sec:CPconserving}
The \CP-conserving observables can be analysed at LO in the large
recoil limit using the heavy-quark and large-$E_{K^*}$ expressions for
the spin amplitudes, as first proposed in \cite{Kruger:2005ep}.  One
of the advantages of this approach is that we obtain analytic
expressions of these observables in a very simple way. These
expressions can be used to study the behaviour of the observables
without having to rely on numerical computations, since the most
relevant features  arise already at LO. The  main goal of this section
is to perform this type of  analysis on the $A_{\mathrm{T}}^{(i)}$
  observables. 

\subsection{Leading-order expressions of \AT2}

 The asymmetry \AT2, first proposed in~\cite{Kruger:2005ep}  is given by
\begin{equation}
  \label{eq:AT2general}
  \AT2 =\frac{|A_{\perp}|^2 - |A_\parallel|^2}{|A_\perp|^2 + |A_\parallel|^2},
\end{equation}
 where $|A_i|^2 = |A_i^L|^2 + |A_i^R|^2$. It has a simple form,  free from
$\xi_{\perp}(0)$ form factor dependencies, in the heavy-quark
($m_{B}\to\infty$) and large  \Kstarzb  energy
($E_{K^*}\to\infty$) limits\footnote{Notice that along this section we
  will drop the superscript ``eff" that \C7 and \C9 should bear in
  order to simplify the notation.}:
\begin{equation}
  \label{eq:AT2generalLEET}
    \AT2 = \frac{2 \left[{\mathrm{Re}}\left(\Cp{10} \C{10}^ {*} \right) + F^2 {\mathrm{Re}}\left(\Cp7 \C7^{*} \right)  
    + F {\mathrm{Re}}\left(\Cp7 \C9^{*} \right)\right]}
    {|\C{10}|^2  +  |\Cp{10}|^2+ F^2 (|\C7|^2 + |\Cp7|^2 ) + |\C9|^2  + 
    2 F {\mathrm{Re}}\left(\C7  \C9^{*} \right)},
\end{equation}
where $F \equiv 2 m_b m_B /q^2$. The Wilson coefficients can take the most general form:
\begin{equation}
\label{eq:WWCCgeneral}
\mathcal{C}_i = \mathcal{C}_i^{\scriptscriptstyle{\mathrm{SM}}}  + 
|\mathcal{C}_i^{\scriptscriptstyle{\mathrm{NP}}}| e^{i \phi_i^{\scriptscriptstyle{\mathrm{NP}}}}, \,\,\,\,\,\,
 \mathcal{C}_{i}^{\prime} = |\mathcal{C}_{i}^\prime| e^{i \phi_{i}^{\prime}}\,,\,\,\,\,    i= 7,9,10.
\end{equation}

We will neglect henceforward both the tiny SM weak phase $\phi_9^{\scriptscriptstyle{\mathrm{SM}}}$, that arises from 
the CKM elements ratio  $\lambda_{u}= ({V_{ub}V_{us}^{*}})/( {V_{tb}V_{ts}^{*}})$, and the SM strong phase 
$\theta_9^{\scriptscriptstyle{\mathrm{SM}}}$, smaller than $1^{{\mathrm{o}}}$ in the low dilepton mass region 
$1\gevgev  \leqslant q^2 \leqslant 6\gevgev$~\cite{Kruger:2000zg}.

Obviously, the observable \AT2 vanishes in the heavy-quark and large
\Kstarzb energy limits at LO when all the Wilson coefficients are
taken to be SM-like.  This result can be understood rather easily. The
left-handed structure of weak interactions in the SM guarantees that,
in these limits, the \s quark created in the $\b \to \s$ transition
will have helicity $h(\s)=-1/2$ in the massless limit ($m_{\s}\to 0$)
\cite{Burdman:2000ku}. This \s quark will combine with the spectator
quark \dbar of the \Bdb to form the \Kstarzb meson
with $h(\Kstarzb)=-1$ or $0$ (but not $+1$), therefore $H_{+}=0$
at quark level in the SM.  Using Eq.~(\ref{hel:trans}), this translates
into $A_{\perp}=-A_{\parallel}$ at the quark level, which corresponds
to $A_{\perp}\simeq-A_{\parallel}$ at the hadron level
\cite{Stech:1995ec,Soares:1996vs,Soares:1998tx}.

\EPSFIGURE[!h]{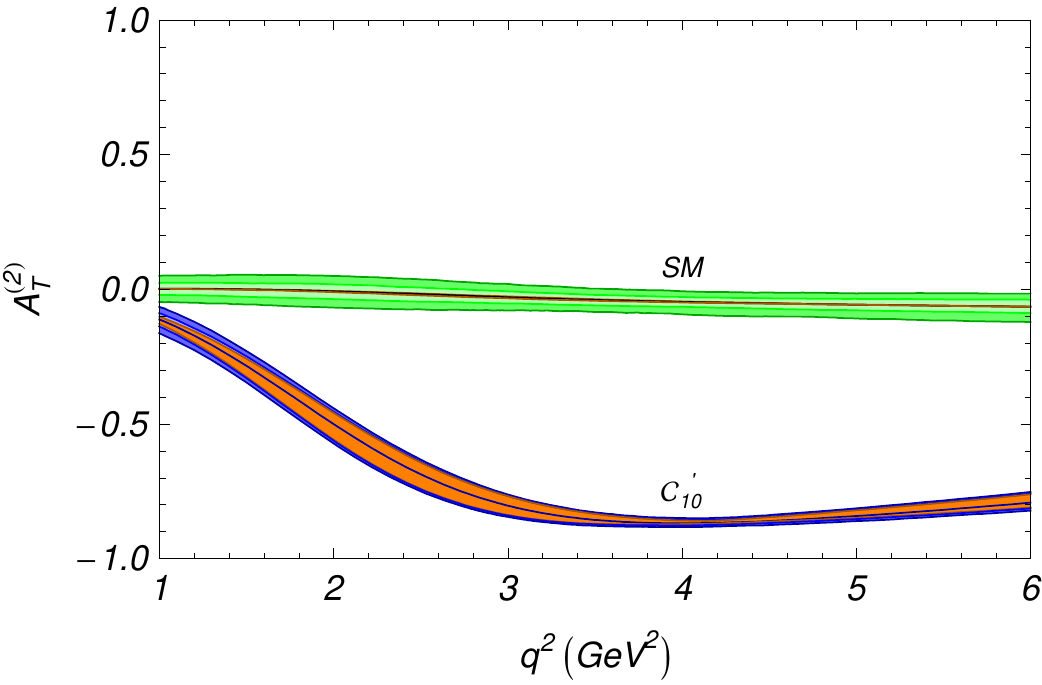,width=0.49\textwidth}{\AT2 in the SM
  (green) and with NP in $\Cp{10}=3 e^{i \frac{\pi}{8}}$ (blue), this
  value is allowed by the model independent analysis of
  \cite{Bobeth:2008ij}.  The inner line corresponds to the central
  value of each curve. The dark orange bands surrounding it are the
  NLO results including all uncertainties (except for $\Lambda_{\rm QCD}/m_b$)
  as explained in the text. Internal light green/blue bands (barely
  visible) include the estimated $\Lambda_{\rm QCD}/m_b$ uncertainty at a $\pm
  5\%$ level and the external dark green/blue bands correspond to a
  $\pm 10\%$ correction for each spin amplitude.
  \label{fig:AT2withNPC10pPi8}
}
The NP dependence of \AT2 can be studied in a model independent way by switching on one Wilson coefficient each 
time and keeping all the others at their SM values. A simple inspection  of Eq.~(\ref{eq:AT2generalLEET}) shows that only
the chirally flipped operators \Opep7 and \Opep{10} give a non-zero expression for 
\AT2 in our approximation:
\begin{equation}
 \label{eq:AT2C7p}
   \AT2 \Big\vert_{7^{\prime}}= \frac{2 F (F \C7^{\scriptscriptstyle{\mathrm{SM}}}+ \C9^{\scriptscriptstyle{\mathrm{SM}}}) |\Cp7| {\mathrm{cos}}(\phi_{7}^{\prime})}
   { (\C{10}^{\scriptscriptstyle{\mathrm{SM}}})^2 + F^2 |\Cp7|^2+ (F \C7^{\scriptscriptstyle{\mathrm{SM}}} + \C9^{\scriptscriptstyle{\mathrm{SM}}})^2} \,,
\end{equation} 
and
 \begin{equation}
 \label{eq:AT2C10p}
   \AT2\Big\vert_{10^{\prime}} = \frac{2 \, \C{10}^{\scriptscriptstyle{\mathrm{SM}}} |\Cp{10}| {\mathrm{cos}}(\phi_{10}^{\prime})}
   {(\C{10}^{\scriptscriptstyle{\mathrm{SM}}})^2 +  |\Cp{10}|^2+(F  \C7^{\scriptscriptstyle{\mathrm{SM}}}  + \C9^{\scriptscriptstyle{\mathrm{SM}}})^2} \, .
\end{equation}
Equations.~(\ref{eq:AT2C10p}) and (\ref{eq:AT2C7p}) show that \AT2 is
sensitive to both the modulus and the sign of the Wilson coefficients
$\Cp7$ and \Cp{10}. When NP enters only $\Cp{10}$, the fact that
$\C{10}<0$ in the SM makes the observable negative unless
$\frac{\pi}{2}< |\phi_{10}^{\prime}| < \pi$, enabling us to
distinguish the sign of this weak phase
(Fig.~\ref{fig:AT2withNPC10pPi8}).  Likewise, if NP appears in $\Cp7$,
\AT2 will display a zero in the dilepton mass spectrum when $F
\C7^{\scriptscriptstyle{\mathrm{SM}}} +
\C9^{\scriptscriptstyle{\mathrm{SM}}}=0$, which will coincide exactly
with the zero of the observable $\AFB$ at LO \cite{Beneke:2001at}. As
the zero is independent of $\Cp7$, all curves with
$\C7^{\scriptscriptstyle{\mathrm{SM}}}$ should exhibit it at $q^2 \sim
4\gevgev$, but if there is also a NP contribution to $\C7$, the zero
will be shifted either to higher or lower values of $q^2$. In case of
a sign flip affecting $\C7$, \AT2 would not have a zero at any value
of $q^2$, exactly as for \AFB (see~\cite{Alok:2009tz}  for a recent discussion of different mechanisms to achieve this). In fact, should NP
enter both \Ope7 and \Opep7 simultaneously,
Eq.~(\ref{eq:AT2generalLEET}) would imply
\begin{equation}
 \label{eq:AT2C7C7p}
   \AT2 \Big\vert_{{7}^{\prime},\,{7}^{\mathrm{NP}}}  \propto 2F \left[(F \C7^{\scriptscriptstyle{\mathrm{SM}}} + 
   \C9^{\scriptscriptstyle{\mathrm{SM}}}) 
   |\Cp7| {\mathrm{cos}}(\phi_{7}^{\prime}) + 
   F  |\Cp7|  |\C7^{\scriptscriptstyle{\mathrm{NP}}}| {\mathrm{cos}}(\phi_{7}^{\prime}-\phi_{7}^{\scriptscriptstyle{\mathrm{NP}}})\right]
\end{equation}
while
\begin{equation}
 \label{eq:AFBC7C7p}
\AFB \Big\vert_{{7}^{\prime},\,{7}^{\mathrm{NP}}} \propto F \C7^{\scriptscriptstyle{\mathrm{SM}}}+ \C9^{\scriptscriptstyle{\mathrm{SM}}} + 
F  |\C7^{\scriptscriptstyle{\mathrm{NP}}}| {\mathrm{cos}}(\phi_{7}^{\scriptscriptstyle{\mathrm{NP}}}). 
\end{equation}
The comparison of Eq.~(\ref{eq:AT2C7C7p}) with Eq.~(\ref{eq:AFBC7C7p})
can be used to explain the improved sensitivity of \AT2 to certain
types of NP versus that of $\AFB$. The numerator of \AT2 exhibits
sensitivity to the weak phases
$\phi_{7}^{\scriptscriptstyle{\mathrm{NP}}}$ and $\phi_{7}^{\prime}$,
having an interference term enhanced by the large factor $F$
($8\lesssim F\lesssim 48$ in the dilepton mass region studied), while
$\AFB$ is only sensitive to
$\phi_{7}^{\scriptscriptstyle{\mathrm{NP}}}$.  Thus, a wider departure
from the SM behaviour is to be expected in \AT2 when NP enters the
operators \Ope7 and \Opep7. This is shown in
Fig.~\ref{fig:AT2andAFBC7C7p} using three different scenarios,
described in the caption of Fig.~\ref{fig:AT2andAFBC7C7p}, compatible
with present experimental and theoretical constraints. Therefore, we
emphasise that \AT2 must be regarded as an improved version of \AFB
once the full-angular analysis becomes possible.  
\FIGURE{
  \includegraphics[width=0.49\textwidth]{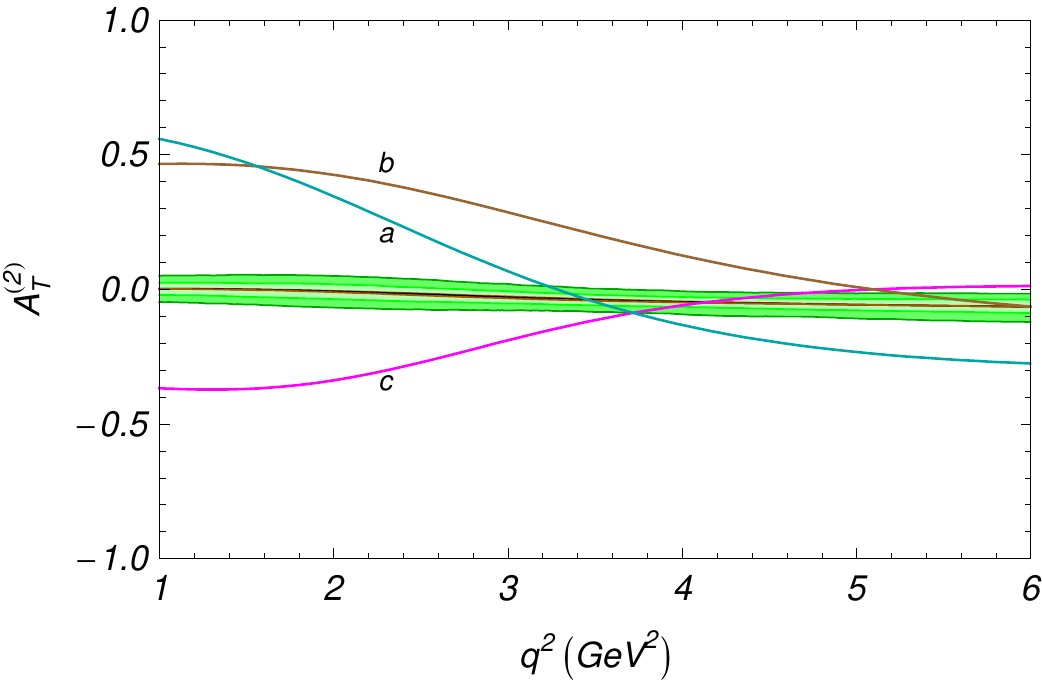}
  \includegraphics[width=0.49\textwidth]{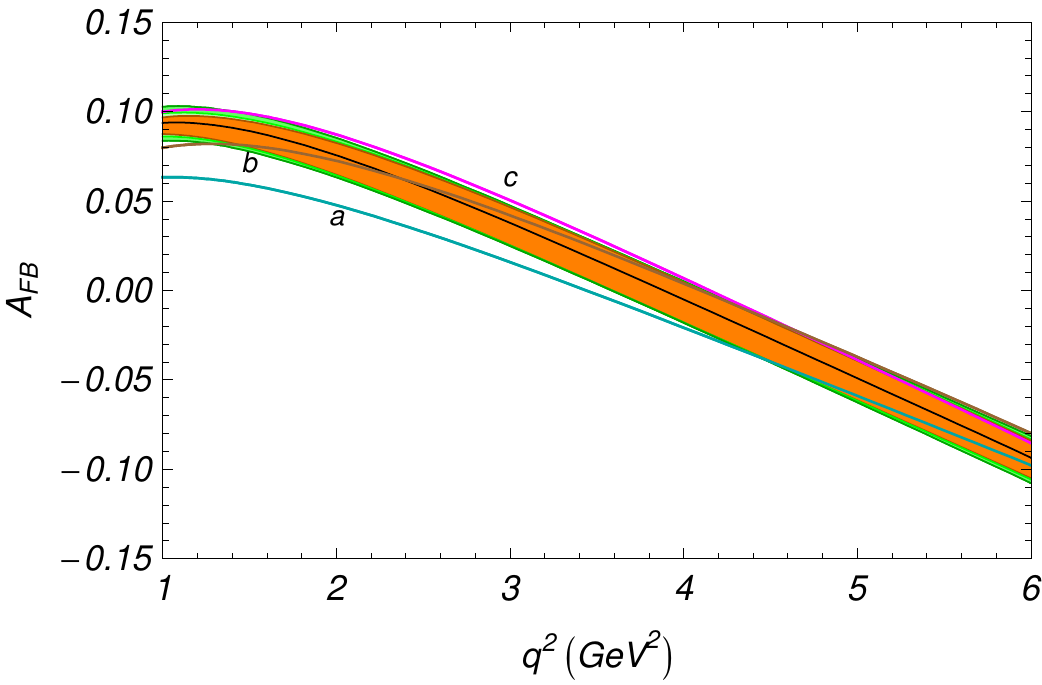}
  \caption{Observables \AT2 and $\AFB$ with NP curves for three
    allowed combinations of $\C7$ and $\Cp7$ following the model
    independent analysis of \cite{Bobeth:2008ij}. The bands correspond
    to the SM and the theoretical uncertainty as described in Fig
    \protect\ref{fig:AT2withNPC10pPi8}. The cyan line (shown with the
    label $a$) corresponds to
    ($\C7^{\scriptscriptstyle{\mathrm{NP}}}$, $\Cp7$) = ($0.26 e^{-i
      \frac{7\pi}{16}}$, $0.2 e^{i \pi}$), the brown line $b$ to
    ($0.07e^{i \frac{3\pi}{5}}$, $0.3 e^{i \frac{3\pi}{5}}$) and the
    magenta line $c$ to ($0.03 e^{i \pi}$, $0.07$).}
  \label{fig:AT2andAFBC7C7p}
}

\subsection{Leading-order expressions of \AT5}

In the SM, we get in the heavy-quark and large-$E_{K^*}$ limits at LO:
\begin{equation}
\label{eq:AT5SMLEET}
  \AT5\Big\vert_{\mathrm{SM}} = \frac{\left| -(\C{10}^{\scriptscriptstyle{\mathrm{SM}}})^2+ (F \C7^{\scriptscriptstyle{\mathrm{SM}}}+ 
  \C9^{\scriptscriptstyle{\mathrm{SM}}})^2 \right|}
  {2 \left[(\C{10}^{\scriptscriptstyle{\mathrm{SM}}})^2 + (F \C7^{\scriptscriptstyle{\mathrm{SM}}}+ \C9^{\scriptscriptstyle{\mathrm{SM}}})^2\right]},
\end{equation}
which sets the ``wave-like'' behaviour of \AT5. At low $q^2$,
Eq.~(\ref{eq:AT5SMLEET}) can be used to check that
$\AT5\big\vert_{SM}^{ 1\gevgev} \!\!\simeq 0.4$. On the other hand, at
the zero-point of \AT2 and $\AFB$, \AT5 exhibits an absolute maximum
of magnitude $\AT5\big\vert_{SM}^{4\gevgev} \!\! \simeq0.5$.

\FIGURE{
  \includegraphics[width=0.49\textwidth]{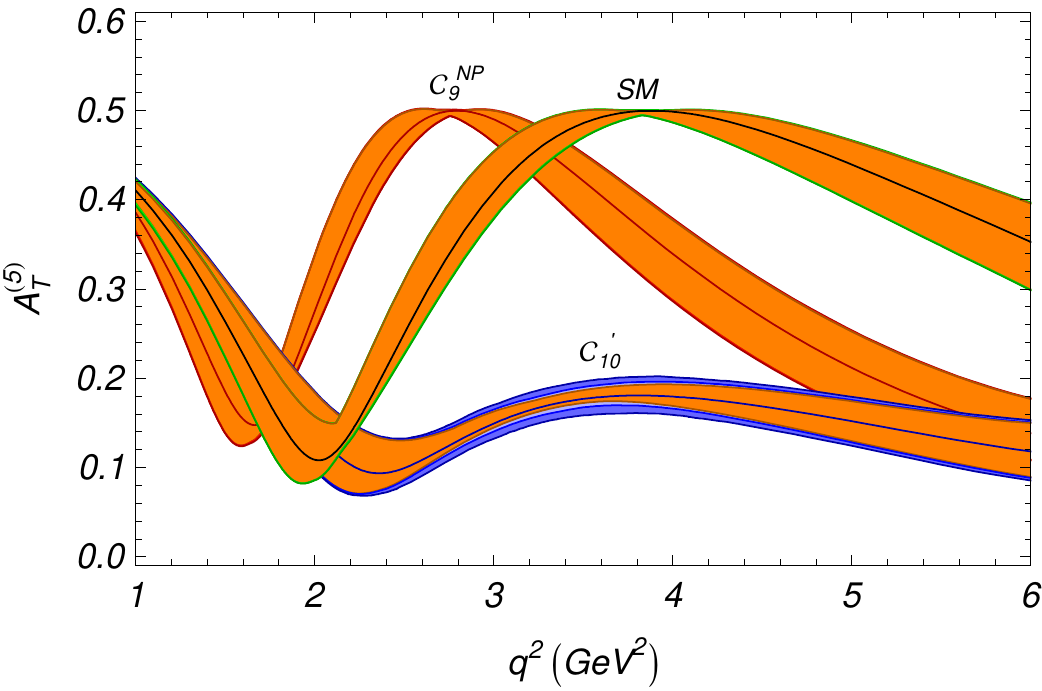}
  \includegraphics[width=0.49\textwidth]{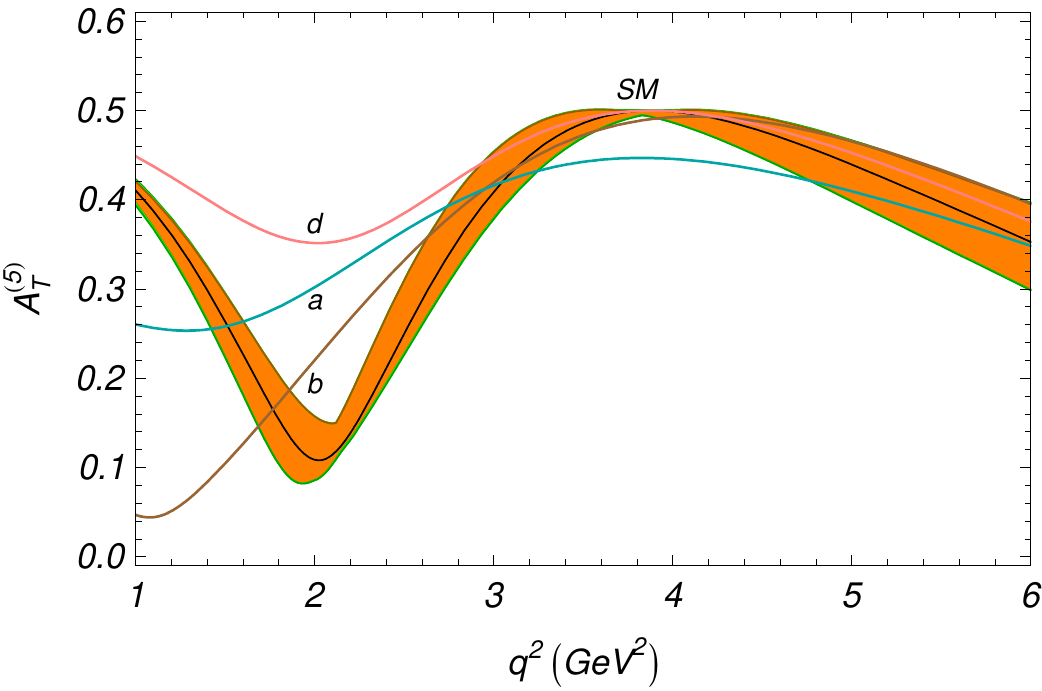}
  \caption{\AT5 in the SM and with NP in $\Cp{10}=3 e^{i
      \frac{\pi}{8}}$ and $\C9^{\scriptscriptstyle{\mathrm{NP}}}=2
    e^{i \frac{\pi}{8}}$ (left) and in both $\C7$ and $\Cp7$ Wilson
    coefficients (right). The cyan line ($a$)
    corresponds to ($\C7^{\scriptscriptstyle{\mathrm{NP}}}$, $\Cp7$) =
    ($0.26 e^{-i \frac{7\pi}{16}}$, $0.2 e^{i \pi}$), the brown line
    ($b$) to ($0.07e^{i \frac{3\pi}{5}}$, $0.3 e^{i \frac{3\pi}{5}}$)
    and the pink line ($d$) to ($0.18 e^{-i \frac{\pi}{2}}$, $0$). The
    bands symbolise the theoretical uncertainty as described in
    Fig.~\protect\ref{fig:AT2withNPC10pPi8}.}
  \label{fig:AT5NP}
}

Any inclusion of NP in the Wilson coefficients $\C7$, $\C9$ and
$\C{10}$ will give rise to the appearance of an extra term in the
numerator (with respect to Eq.~(\ref{eq:AT5SMLEET})) that will shift
the observable along the $y$-axis.
\begin{subequations}
  \label{eq:AT5NP}
  \begin{align}
  \label{eq:AT5C7NPpi2}
   \AT5 \Big\vert_{7^{\mathrm{NP}}}^{\pi/2} &= \frac{\sqrt{\left[-(\C{10}^{{\scriptscriptstyle{\mathrm{SM}}}})^2 + 
   F^2 |\C7^{\scriptscriptstyle{\mathrm{NP}}}|^2 + 
   (F \C7^{\scriptscriptstyle{\mathrm{SM}}} + \C9^{\scriptscriptstyle{\mathrm{SM}}})^2\right]^2 + 
   4 \left[F \C{10}^{\scriptscriptstyle{\mathrm{SM}}}  |\C7^{\scriptscriptstyle{\mathrm{NP}}}|\right]^2}}
   {2 \left[ (\C{10}^{{\scriptscriptstyle{\mathrm{SM}}}})^2 + F^2 |\C7^{\scriptscriptstyle{\mathrm{NP}}}|^2 + 
   (F \C7^{\scriptscriptstyle{\mathrm{SM}}} + \C9^{\scriptscriptstyle{\mathrm{SM}}})^2\right]}\,, \\
  \label{eq:AT5C9NPpi2}
   \AT5\Big\vert_{9^{\mathrm{NP}}}^{\pi/2} &= \frac{\sqrt{\left[-(\C{10}^{{\scriptscriptstyle{\mathrm{SM}}}})^2 + 
   |\C9^{\scriptscriptstyle{\mathrm{NP}}}|^2 + (F \C7^{\scriptscriptstyle{\mathrm{SM}}} + 
   \C9^{\scriptscriptstyle{\mathrm{SM}}})^2 \right]^2 + 
   4 \left[\C{10}^{\scriptscriptstyle{\mathrm{SM}}} |\C9^{\scriptscriptstyle{\mathrm{NP}}}|\right]^2}}
   {2 \left[(\C{10}^{{\scriptscriptstyle{\mathrm{SM}}}})^2 + |\C9^{\scriptscriptstyle{\mathrm{NP}}}|^2 + 
   (F \C7^{\scriptscriptstyle{\mathrm{SM}}}+\C9^{\scriptscriptstyle{\mathrm{SM}}})^2\right]}
\,, \\
  \label{eq:AT5C10NPpi2}
   \AT5 \Big\vert_{10^{\mathrm{NP}}}^{\pi/2} &= \frac{\sqrt{\left[-(\C{10}^{{\scriptscriptstyle{\mathrm{SM}}}})^2 -
    |\C{10}^{\scriptscriptstyle{\mathrm{NP}}}|^2 + 
   (F \C7^{\scriptscriptstyle{\mathrm{SM}}} + \C9^{\scriptscriptstyle{\mathrm{SM}}})^2\right]^2 + 
   4 \left[|\C{10}^{\scriptscriptstyle{\mathrm{NP}}}| (F \C7^{\scriptscriptstyle{\mathrm{SM}}} + \C9^{\scriptscriptstyle{\mathrm{SM}}})\right]^2}}
   {2 \left[ (\C{10}^{{\scriptscriptstyle{\mathrm{SM}}}})^2  + |\C{10}^{\scriptscriptstyle{\mathrm{NP}}}|^2 + 
   (F \C7^{\scriptscriptstyle{\mathrm{SM}}} + \C9^{\scriptscriptstyle{\mathrm{SM}}})^2\right]} \, .
\end{align}
\end{subequations} 
In Eq.~(\ref{eq:AT5NP}) we have chosen
for simplicity the weak phase
$\phi_i^{\scriptscriptstyle{\mathrm{NP}}}=\pi/2$ for $i=7,\,9,\,10$,
but they turn out to be dominated by the SM contribution unless the NP
Wilson coefficients are very large. However, if the weak phases
associated to NP Wilson coefficients are different from $\pi/2$, the
\AT5 curve will get shifted either to the left or to the right,
depending on the value of the angle, as shown in Fig.~\ref{fig:AT5NP}.
 
NP might also enter via the chirally flipped \Opep7 and \Opep{10}. 
The corresponding LO expressions of \AT5 in the heavy-quark and high-$E_{K^*}$ limits read 
\begin{equation}
  \label{eq:AT5C7pLEET}
    \AT5\Big\vert_{7^{\prime}} = \frac{\left|-  (\C{10}^{{\scriptscriptstyle{\mathrm{SM}}}})^2  + (F \C7^{\scriptscriptstyle{\mathrm{SM}}}+
    \C9^{\scriptscriptstyle{\mathrm{SM}}})^2 - F^2 |\Cp7|^2 \right|}
    {2 \left[ (\C{10}^{{\scriptscriptstyle{\mathrm{SM}}}})^2  + (F \C7^{\scriptscriptstyle{\mathrm{SM}}}+\C9^{\scriptscriptstyle{\mathrm{SM}}})^2 + F^2 |\Cp7|^2\right] }
\end{equation}
and
\begin{equation}
  \label{eq:AT5C10pLEET}
  \AT5\Big\vert_{10^{\prime}} = \frac{\left| -  (\C{10}^{{\scriptscriptstyle{\mathrm{SM}}}})^2  +
      |\Cp{10}|^2 + (F \C7^{\scriptscriptstyle{\mathrm{SM}}} + 
      \C9^{\scriptscriptstyle{\mathrm{SM}}})^2 \right|}
  {2 \left[ (\C{10}^{{\scriptscriptstyle{\mathrm{SM}}}})^2  + |\Cp{10}|^2 + (F \C7^{\scriptscriptstyle{\mathrm{SM}}} + \C9^{\scriptscriptstyle{\mathrm{SM}}})^2\right]}.
\end{equation}
Equations~(\ref{eq:AT5C7pLEET}) and (\ref{eq:AT5C10pLEET}) are both free
from NP weak-phase dependence. \AT5 evaluated at the \qsq value of the $\AFB$ zero-point can be
computed easily using Eq.~(\ref{eq:AT5C10pLEET}), obtaining
\begin{equation}
  \AT5\big\vert_{q^2_0} = \frac{1}{2}\frac{|-(\C{10}^{\scriptscriptstyle{\mathrm{SM}}})^2+|\Cp{10}|^2|}{(\C{10}^{\scriptscriptstyle{\mathrm{SM}}})^2+|\Cp{10}|^2},
\end{equation}
where the choice $\Cp{10}=0$ enables us to recover the SM prediction
$\AT5\big\vert_{SM}^{4\gevgev} \!\! = 0.5$.  In Fig.~\ref{fig:AT5NP}
(left) it can be seen that for $|\Cp{10}|=3$ the departure of the NP
curve obtained from the SM behaviour is indeed large.

\subsection{Analysis of \AT3 and \AT4}
The observables \AT3 and \AT4 were first introduced in~\cite{Egede:2008uy} to test the longitudinal spin amplitude $A_{0}$ in a controlled way:
\begin{equation}
 \label{eq:AT3andAT4}
    \AT3 =\frac{|A_{0L} A_{\parallel L}^* + A_{0R}^* A_{\parallel R}|}{\sqrt{|A_0|^2 |A_\perp|^2}}, \qquad\qquad
    A_T^{{(4)}} =\frac{ |A_{0L} A_{\perp L}^* - A_{0R}^* A_{\perp R}|}{|A_{0L} A_{\parallel L}^*+A_{0R}^* A_{\parallel R}|}.
\end{equation}
Unfortunately, the simultaneous appearance of $A_{\perp}$,
$A_{\parallel}$ and $A_{0}$ inside square roots turns the heavy-quark
and large-energy limits into rather awkward expressions, not really
useful to explain the behaviour of these observables at a
glance. Therefore, we only outline their general properties.
Equation~(\ref{eq:AT3andAT4}) shows that \AT3 and \AT4 play a
complementary role, as the numerator of \AT3 and the denominator of
\AT4 are the same. Thus, when a minimum appears in one of them, a
maximum is expected in the other observable and the other way
around. This is indeed what can be observed in
Fig.~\ref{fig:AT3andAT4NP}. For the values of the Wilson coefficients
chosen, NP entering $\Cp{10}$ can easily be distinguished from the SM
curve, displaying a maximum at around 3.5-4\gevgev (exactly in the
energy region where \AT4 is showing a minimum), while
$\C{10}^{\scriptscriptstyle{\mathrm{NP}}}$ can only be clearly
identified using \AT4. Something similar happens with NP entering
$\C7^{\scriptscriptstyle{\mathrm{SM}}}$ and $\Cp7$: the
model-independent values chosen for these Wilson coefficients do not
give rise to clear NP signals from \AT3, but they can be easily told
apart using \AT4. In those situations where the origin of the NP curve
can not be clearly established using a single observable (for
instance, the $c$ curve in the \AT4 plot of Fig.~\ref{fig:AT3andAT4NP}
is very similar to the $\C{10}^{\scriptscriptstyle{\mathrm{NP}}}$
curve), the combined use of \AT2, \AT3, \AT4, \AT5 and maybe $\AFB$
enables us to identify which Wilson coefficient(s) has a contribution
from NP.
\FIGURE{
  \includegraphics[width=0.49\textwidth]{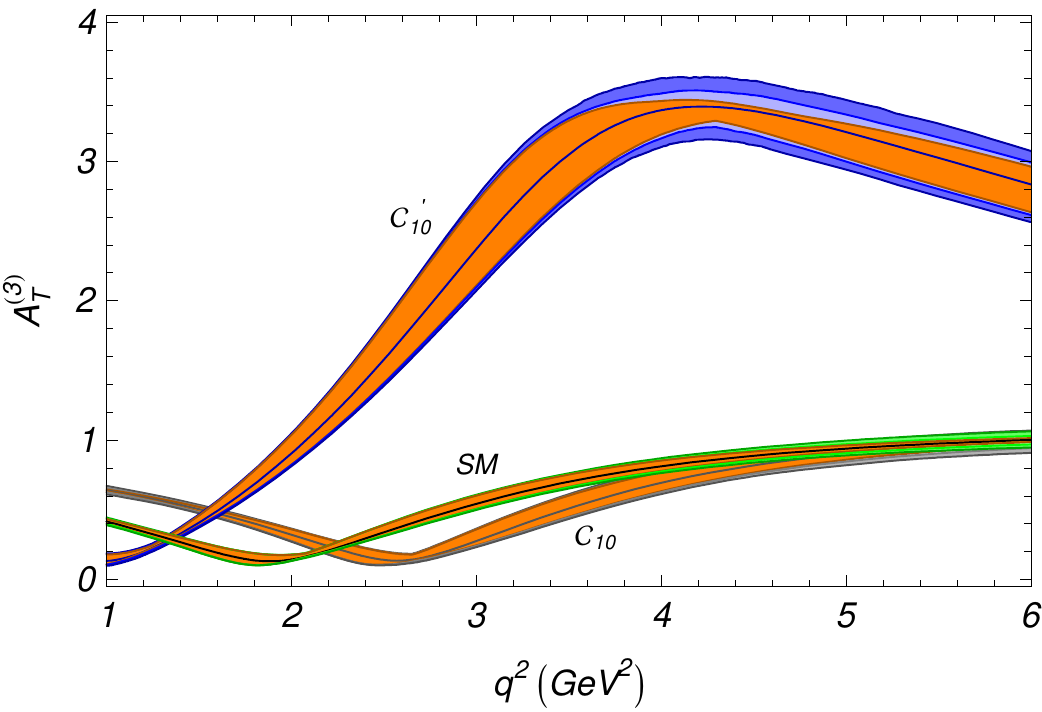}
  \includegraphics[width=0.49\textwidth]{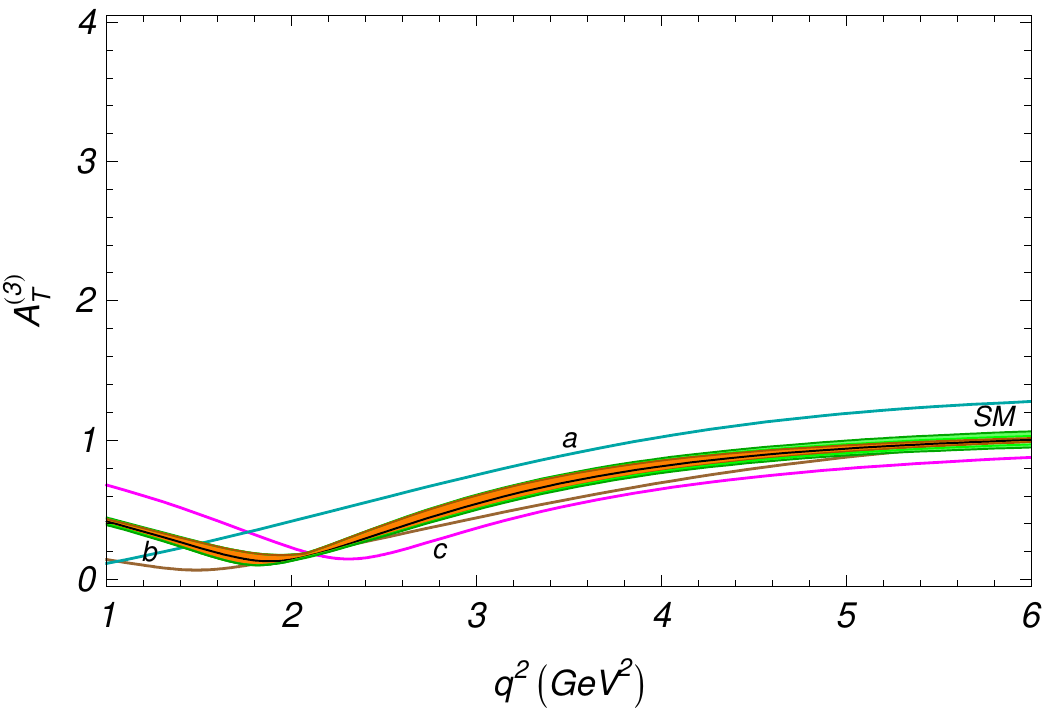}\\ 
  \includegraphics[width=0.49\textwidth]{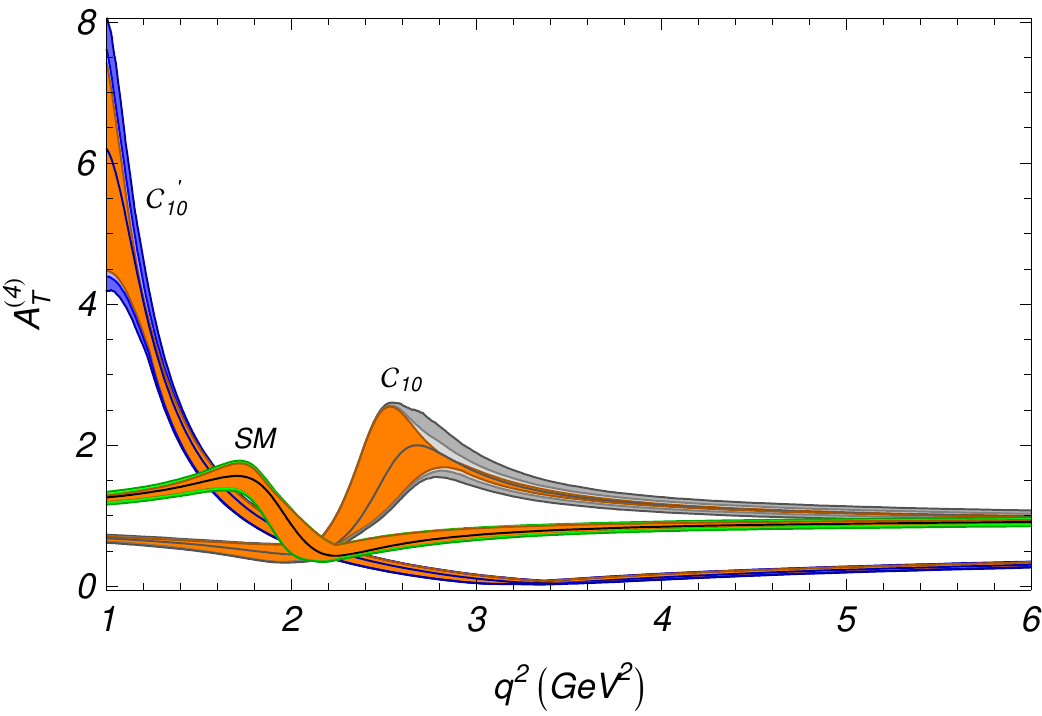}
  \includegraphics[width=0.49\textwidth]{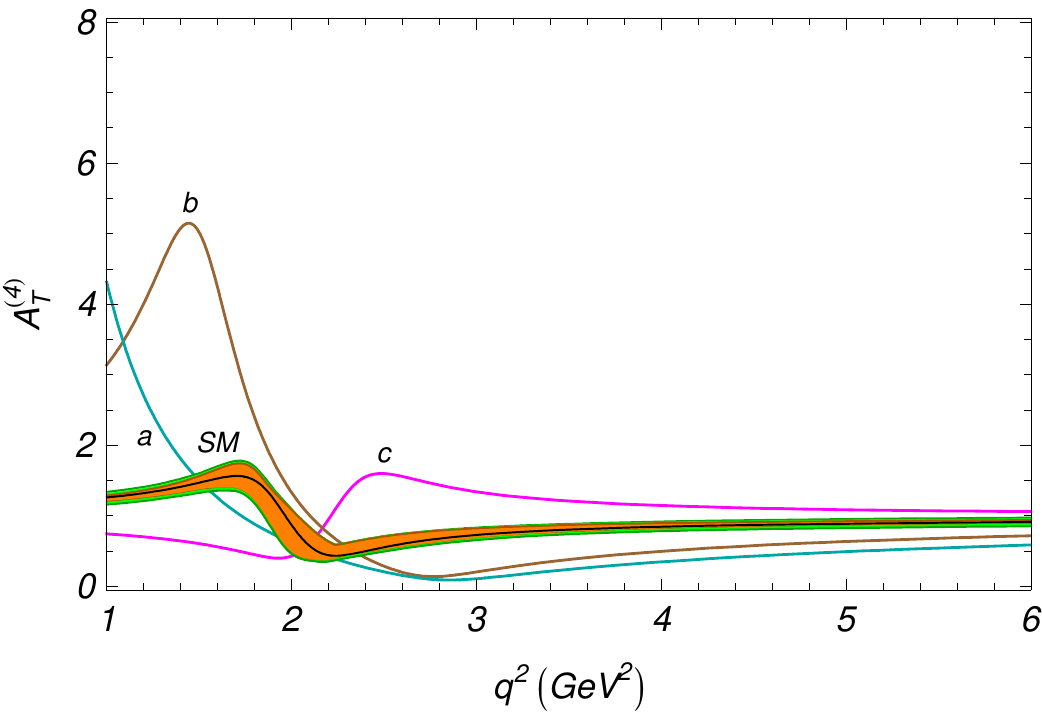}
  \caption{\AT3 and \AT4 in the SM and with NP in and
    $\C{10}^{\scriptscriptstyle{\mathrm{NP}}}=1.5 e^{i \frac{\pi}{8}}$
    and $\Cp{10}=3 e^{i \frac{\pi}{8}}$ (left) and in both $\C7$ and
    $\Cp7$ Wilson coefficients (right).  The cyan line (curve $a$)
    corresponds to ($\C7^{\scriptscriptstyle{\mathrm{NP}}}$, $\Cp7$) =
    ($0.26 e^{-i \frac{7\pi}{16}}$, $0.2 e^{i \pi}$), the brown line
    (curve $b$) to ($0.07e^{i \frac{3\pi}{5}}$, $0.3 e^{i \frac{3\pi}{5}}$)
    and the magenta line (curve $c$) to ($0.03 e^{i \pi}$, $0.07$).
    The bands symbolise the theoretical uncertainty as described
    in Fig.~\protect\ref{fig:AT2withNPC10pPi8}.}
  \label{fig:AT3andAT4NP}
}

\section{Conclusion}
In this paper we have presented how the decay \BdbKsll can provide
detailed knowledge of NP effects in the flavour sector. We developed a
method for constructing observables  with specific sensitivity to
  some types of  NP  while, at the same time,
 keeping theoretical errors from form factors under
control.  A method based on infinitesimal symmetries was
presented which allows in a generic way to identify if an arbitrary
combination of spin amplitudes is an observable of the angular
distribution. For the case of massless leptons we identified the
explicit form of all four symmetries present.   We showed
the possible impact of the unknown $\Lambda_{\rm QCD}/m_b$ corrections on the
NP sensitivity of the various angular observables in a
systematic way using an ensemble method.  Experimental
sensitivity to the observables was evaluated for datasets
corresponding to 10\invfb of data at \lhcb. Using these tools, we did
a phenomenological analysis for both \CP-conserving and \CP-violating
observables.  The  conclusion from this is that the
\CP-violating observables have very poor experimental sensitivity
while the \CP-conserving observables $A_{\mathrm{T}}^{(i)}$ (with
$i=2,3,4$) are very powerful for finding NP,  including
situations  with large weak phases.

\enlargethispage{\baselineskip}

\acknowledgments 
JM acknowledges financial support from FPA2005-02211, 2005-SGR-00994,
MR from the Universitat Aut\`onoma de Barcelona, UE and WR from the
Science and Technology Facilities Council~(STFC), and TH from the
European network Heptools. TH thanks the CERN theory group for its
hospitality during his visits to CERN.

\newpage 

\appendix

\section{Kinematics}
\label{App:Kinematics}
Assuming the \Kstarzb to be on the mass shell, the decay $\BdbKsll$ is
completely described by four independent kinematic variables; namely,
the square of the lepton-pair invariant mass, \qsq, and the three
angles $\theta_l$, $\theta_{K}$ and $\phi$. The sign of the angles for
the \Bdb decay shows great variation in the literature. Therefore we
present here  an  explicit definition of our conventions and point
out where the same or different definitions have been used.

First we consider the $\BdbKsll$ decay. The angle $\theta_l$ is the
angle between the \mup momentum in the rest frame of the dimuon and
the direction of the dimuon in the rest frame of the \Bdb. The
$\theta_K$ angle is in a similar way the angle between the \Km momentum
in the \Kstarzb rest frame and the direction of the \Kstarzb in the
rest frame of the \Bdb.

Let us for $\BdbKsll$ define the momentum vectors
\begin{eqnarray}
  \label{eq:VectorDefs}
  \vec{P}_{\ellp\ellm} & = & \vec{p}_{\ellp} + \vec{p}_{\ellm} \,,\\
  \vec{Q}_{\ellp\ellm} & = & \vec{p}_{\ellp} - \vec{p}_{\ellm} \,,\\
  \vec{P}_{\Km\pip} & = & \vec{p}_{\Km} + \vec{p}_{\pip}   \,, \\
  \vec{Q}_{\Km\pip} & = & \vec{p}_{\Km} - \vec{p}_{\pip} \, .
\end{eqnarray}
In the dimuon rest frame, we have that the \ellp momentum is parallel
to $\vec{Q}_{\ellp\ellm}$ and also that $\vec{P}_{\Km\pip}$ points in
the opposite direction of the dimuon in the \Bdb rest frame. Thus we
can compute the $\theta_l$ angle as
\begin{equation}
  \label{eq:ThetaLDef}
  \cos \theta_l  =  - \frac{\vec{Q}_{\ellp\ellm}^{\ell\ell} \cdot \vec{P}_{\Km\pip}^{\ell\ell}}
                            {|\vec{Q}_{\ellp\ellm}^{\ell\ell}|  |\vec{P}_{\Km\pip}^{\ell\ell}|} \, ,
\end{equation}
where the superscript is used to indicate the frame. In a similar way
we have in the \Kstarzb rest frame
\begin{equation}
  \label{eq:ThetaKDef}
  \cos \theta_K =  - \frac{\vec{Q}_{\Km\pip}^{\Kstar} \cdot \vec{P}_{\ellp\ellm}^{\Kstar}}
                           {|\vec{Q}_{\Km\pip}^{\Kstar}|  |\vec{P}_{\ellp\ellm}^{\Kstar}|}  \, .
\end{equation}
Finally, if we go to the rest frame of the \Bdb, we have $\phi$ as the
signed angle between the planes defined by the two muons and the
\Kstarzb decay products respectively. Vectors perpendicular to the
decay planes are
\begin{equation}
  \label{eq:PerpVectorDef}
  \vec{N}_{\ellp\ellm} = \vec{P}_{\ellp\ellm}^{\B} \times \vec{Q}_{\ellp\ellm}^{\B} \,, \qquad 
  \vec{N}_{\Km\pip} = \vec{P}_{\Km\pip}^{\B} \times \vec{Q}_{\Km\pip}^{\B} \,,
\end{equation}
which lets us define $\phi$ from
\begin{equation}
  \label{eq:PhiDef}
  \cos \phi = - \frac{\vec{N}_{\ellp\ellm} \cdot \vec{N}_{\Km\pip}}{|\vec{N}_{\ellp\ellm}| |\vec{N}_{\Km\pip}|} \,,
  \qquad
  \sin\phi = \left(\frac{\vec{N}_{\ellp\ellm} \times \vec{N}_{\Km\pip}}
                          {|\vec{N}_{\ellp\ellm}| |\vec{N}_{\Km\pip}|}
               \right) \cdot
               \frac{\vec{P}_{\ellp\ellm}^{\B}}{|\vec{P}_{\ellp\ellm}^{\B}|} \, .
\end{equation}
The angles are defined in the intervals
\begin{equation}
  -1\leqslant\cos\theta_l\leqslant 1\, ,\qquad
  -1\leqslant\cos\theta_{K}\leqslant 1\, , \qquad
  -\pi\leqslant\phi < \pi\, .
\end{equation}
 The definition  given here is identical to~\cite{Egede:2008uy}
but is different to~\cite{Altmannshofer:2008dz}. However, the two
definitions result in the same signs for all the coefficients $J_i$
in Eq.~(\ref{eq:AC-last}).

Now for the \BdKsll decay the $\theta_l$ angle is still specified with
respect to the \ellp while for $\theta_K$ the angle is for the
\Kp. This is equivalent to what is done
in~\cite{Altmannshofer:2008dz}. As the $\theta_l$ angle does not
change the sign of the lepton, we have
\begin{equation}
  \label{eq:CPVangleDef}
  \bar{J}_{1,2,3,4,7} = J_{1,2,3,4,7} \,, \qquad \bar{J}_{5,6,8,9} = - J_{5,6,8,9} \, .
\end{equation}
in the full-angular distribution in the absence of \CP violation.

For the experimental  papers~\cite{:2009zv,Aubert:2006vb}, a definition
has been adopted where all angular distributions have been plotted for
the \BdKsll decay, with the \BdbKsll events overlaid assuming \CP
conservation. In practise this means that \BdbKsll events have the
sign of $\cos\theta_l$ reversed before plotting. When experiments
progress to measuring the $\phi$ angle as well, special care needs to
be taken to get the definitions correct.

\section{Theoretical input parameters and uncertainties}
\label{sec:InputAppendix}
To compute the soft form factor error bands in
Figs.~\ref{formfactorA62s} and~\ref{formfactorA8} in a conservative
fashion, we have used, as input data, the values of $\xi_\parallel(0)$
and $\xi_\perp(0)$ shown in Table~\ref{App:Input}. One can notice that
the $\xi_\perp(0)$ value has been taken from \cite{Altmannshofer:2008dz}, as it is compatible to 
$\xi_\perp(0)=0.26$ used in \cite{Beneke:2004dp}, while for $\xi_\parallel(0)$ we have kept
the value from \cite{Beneke:2004dp} to allow for a wider uncertainty
range.

The $q^2$-dependence of the form factors $V$, $A_1$ and $A_2$ has been
parametrised according to \cite{Ball:1998kk}
\begin{equation}
\label{eq:fiteqVA1A2}
 F(q^2)=\frac{F(0)}{1-a_F q^2/m_B^2 + b_F q^4/m_B^4},
\end{equation}
where $F(0)$, $a_F$ and $b_F$ are the fit parameters shown in Table 3
of \cite{Ball:1998kk}. Substituting the outcomes of
Eq.~(\ref{eq:fiteqVA1A2}) into~\cite{Beneke:2004dp}
\begin{eqnarray}
\label{eq:xisBeneke2}
 \xi_\perp(q^2)\!\!&=&\!\!\frac{m_B}{m_B+m_{K^*}} V(q^2) \,, \nonumber\\
 \xi_\parallel(q^2)\!\!&=&\!\!\frac{m_B+m_{K^*}}{2E_{K^*}} A_1(q^2)-\frac{m_B-m_{K^*}}{m_B}A_2(q^2) \,,
\end{eqnarray}
we can obtain both the central value and the associated uncertainty curves 
for $\xi_\parallel(q^2)$ and $\xi_\perp(q^2)$ in the $1$-$6\gevgev$
range. These are used to get the fitting parameters $A$, $B$, $C$ and $D$ of
\begin{eqnarray}
\label{eq:xisBeneke1}
 \xi_\perp(q^2)\!\!&=&\!\!\xi_{\perp}(0)\left(\frac{1}{A - B (q^2/m_{B}^2)}\right)^2  \,, \nonumber\\
 \xi_\parallel(q^2)\!\!&=&\!\!\xi_{\parallel}(0)\left(\frac{1}{C - D (q^2/m_{B}^2)}\right)^3  \,,
\end{eqnarray}
where $A,C \simeq 1$ within a per mille precision.
This parametrisation follows closely Eq.~(47) in \cite{Beneke:2001at} and allows us to explore the impact of 
$\xi_{\perp}(0)$ and $\xi_{\parallel}(0)$ (with their corresponding uncertainties) to the $CP$-violating and 
$CP$-conserving observables studied throughout this paper.

The next step is to compute the amplitudes, keeping one soft form
factor fixed at the central value and varying the other in the range
allowed by its uncertainty. From them, the observables can be obtained
in a straightforward way and the errors added in quadrature.

To generate the theoretical error bands not due to $\Lambda/m_b$
corrections (plotted as the inner orange strips in the plots of
Secs.~\ref{sec:CPviolating} and~\ref{sec:CPconserving}) we have used
the criteria of Beneke et al. in \cite{Beneke:2001at} and added the
following uncertainties in quadrature: the renormalisation scale
uncertainty has been found by varying $\mu$ between $2.3$ and
$9.2\gev$ (where $\mu$ is the scale at which the Wilson Coefficients,
$\alpha_s$ and the $\overline{\mathrm{MS}}$ masses are evaluated), the
uncertainty in the ratio $m_c/m_b$ by varying this quantity between
0.29 and 0.31, and the other parametric uncertainties have been
collected into the factor \cite{Egede:2008uy}
\begin{equation}
  \kappa(q^2)=\frac{\pi^2 f_B f_{K^*,z}(\mu)}{N_c m_B \xi_z(q^2)} \qquad {\mathrm{with}}\,\,z=\perp,\parallel
\end{equation}
that determines the relative magnitude of the hard-scattering versus the form factor term \cite{Beneke:2001at}, 
which is uncertain by about $\pm 35\%$.  In our numerical analysis  we have used the values of the Wilson coefficients in Table~1 of Ref.~\cite{Beneke:2001at}.

\TABLE{
  \centerline{\parbox{14cm}{\caption{\label{App:Input}
        Summary of input parameters and estimated uncertainties.}}}
  \vspace{-0.1cm}
    \begin{tabular}{| l l| l l |} 

\hline 
\hline  
\rule[-2mm]{0mm}{7mm}
     $\!\!m_B$ \cite{Amsler:2008zzb}       &$5279.50 \pm 0.30\mev$ & 
     	$\lambda$ \cite{Amsler:2008zzb}      & $0.226 \pm 0.001$ \\
     
     $m_K$ \cite{Amsler:2008zzb}      &$896.00 \pm 0.25\mev$ & 
         $A$ \cite{Amsler:2008zzb}      & $0.814 \pm 0.022$ \\
	
     $M_W$ \cite{Amsler:2008zzb}      & $80.398\pm 0.025\gev$  & 
        $\bar\rho$ \cite{Amsler:2008zzb}      & $0.135 \pm 0.031$ \\
        
     $M_Z$	 \cite{Amsler:2008zzb}      &$91.1876 \pm 0.0021\gev$ &
       $\bar\eta$ \cite{Amsler:2008zzb}      & $0.349 \pm 0.017$
        
   \\[0.15cm]
\hline
\rule[-2mm]{0mm}{7mm}
     $\!\!\hat m_t(\hat m_t)$ \cite{Beneke:2001at}           & $167 \pm 5\gev$ &
      $\Lambda_{\rm QCD}^{(n_f=5)}$ \cite{Amsler:2008zzb}     & $220 \pm 40\mev$\\
    
     $\hat m_b(\hat m_b)$ \cite{Buchmuller:2005zv}      & $4.20\pm 0.04\gev$ &
      $\alpha_{s}(M_Z)$ \cite{Amsler:2008zzb}      & $0.1176\pm 0.0002$ \\

     $\hat m_c(\hat m_c)$ \cite{Zyablyuk:2002kg}     & $1.26\pm0.02 \gev$ & 
      $\alpha_{\rm em}$ \cite{Amsler:2008zzb}      & $ 1/137$

\\[0.15cm]
\hline
\rule[-2mm]{0mm}{7mm}
     $\!\!f_B$ \cite{Onogi:2006km}         & $200 \pm 25\mev$ &
      $a_{1,\,K^*}^{\perp,\,\parallel}(2\gev)$ \cite{Ball:2007zt}         &  $0.03 \pm 0.03$\\
   
     $f_{K^*}^{\perp}(2\gev)$ \cite{Ball:2007zt}         & $163 \pm 8\mev$ &
      $a_{2,\,K^*}^{\perp,\,\parallel}(2\gev)$ \cite{Ball:2007zt}        &  $0.08 \pm 0.06$\\
        
     $f_{K^*}^{\parallel}(2\gev)$ \cite{Ball:2007zt}             & $220 \pm 5\mev$  &&

\\[0.15cm]
\hline
\rule[-2mm]{0mm}{7mm}
     $\!\!m_B\,\xi_{\parallel}(0)/(2m_{K^*})$ \cite{Beneke:2004dp}         & $ 0.47 \pm 0.09$ &
       	$\lambda_{B,+} (\mu_h)$ \cite{Ball:2006nr}          & $0.51 \pm 0.12\gev$ \\
     $\xi_{\perp}(0)$ \cite{Altmannshofer:2008dz}             & $  0.266 \pm 0.032$ &
        $\mu_h$ \cite{Altmannshofer:2008dz}            & $  2.2\gev$
\\[0.15cm]
\hline
\hline  
\end{tabular} 
}

\newpage 

\bibliographystyle{JHEP}
\bibliography{paper}

\end{document}